\documentclass[12pt]{article}
\usepackage{amsfonts}

\usepackage{graphicx}
\usepackage{amsmath}
\usepackage{float}

\textwidth=6.5in \hoffset=-.75in \voffset=-1in \numberwithin{equation}{section}
\numberwithin{figure}{section}

\input{tcilatex}

\begin{document}

\begin{titlepage}
\bigskip \begin{flushright}
hep-th/0404071\\
\end{flushright}
\vspace{1cm}
\begin{center}
{\Large \bf {New Reducible Membrane Solutions in $D=11$ Supergravity}}\\
\end{center}
\vspace{2cm}
\begin{center}
R. Clarkson{ \footnote{ EMail: r2clarks@sciborg.uwaterloo.ca}} , A.M.
Ghezelbash{ \footnote{ EMail: amasoud@avatar.uwaterloo.ca}}, R.B. Mann{
\footnote{ EMail: mann@avatar.uwaterloo.ca}}\\
Department of Physics, University of Waterloo, \\
Waterloo, Ontario N2L 3G1, Canada\\
\vspace{1cm}
\today\\
\end{center}

\begin{abstract}
We construct generalized 11D supergravity solutions of fully localized D2/D6 brane
intersections. These solutions are obtained by embedding Taub-NUT and/or
self-dual geometries lifted to M-theory. We reduce these
solutions to ten dimensions, obtaining new D-brane systems in type IIA supergravity.
We discuss the limits in which the
dynamics of the D2 brane decouples from the bulk for these solutions.
\end{abstract}
\end{titlepage}\onecolumn 
\bigskip 

\section{Introduction}

It is believed that fundamental M-theory in the low-energy limit can be
described effectively by $D=11$ supergravity \cite{gr1}. Extending our
understanding of the different classical brane solutions in string theory
(or M-theory) is important, and so there is a lot of interest in finding $%
D=11$ M-brane solutions such that after reduction to ten dimensions, they
(or some combinations of them) reduce simply to the supersymmetric BPS
saturated $p$-brane solutions. Some supersymmetric solutions of two or three
orthogonally intersecting 2-branes and 5-branes in $D=11$\ supergravity were
obtained some years ago \cite{Tsey}, and more such solutions have since been
found \cite{other,smith}.

Recently an interesting new supergravity solution for a localized D2/D6
intersecting brane system was found \cite{Hashi}. The solution was
constructed by lifting a D6 brane to a four-dimensional Taub-NUT geometry
embedded in M-theory and then placing M2-branes in the Taub-NUT background
geometry.

A special feature of this construction\ is that the solution is not
restricted to be in the near core region of the D6 brane. Moreover, in \cite%
{Ali}, partially localized D-brane systems involving D3, D4 and D5 branes
were constructed.{\large \ }By assuming a simple ansatz for the eleven
dimensional metric, the problem reduces to a partial differential equation
that is separable and admits proper boundary conditions.{\large \ }

Motivated by this work, we consider extensions of these solutions to more
general types of Taub-NUT and/or self-dual geometries lifted to M-theory. We
have obtained several different solutions of interest. Specifically, since
in the 11 dimensional metric for an M2 brane, the M2 brane itself only takes
up two of the 10 spatial coordinates, we can embed a variety of different
geometries. \ These include the Eguchi-Hanson metric, higher-dimensional
forms of the Taub-NUT metric, and products of these 4-dimensional metrics.
After compactification on a circle, we find the different fields of type IIA
string theory.

In our procedure we begin with a general ansatz for the metric function of
an M2 brane in 11-dimensional M-theory. After compactification on a circle $%
(T^{1}),$\ we find a solution to type IIA theory for which the highest
degree of the field strengths is four. \ Hence the non-compact global
symmetry for massless modes is given by the maximal symmetry group $E_{1(1)}=%
\mathbb{R}$, without any need to dualize the field strengths \cite{cremmer}.
For the full type IIA theory, only the discrete subgroup $E_{1(1)}(\mathbb{Z}%
)=\mathbb{Z}$ survives, in particular by its action on the BPS spectrum and
as a discrete set of identifications on the supergravity moduli space. This
subgroup is the U-duality group for all type IIA theories we find in this
paper.{\large \ }

In section \ref{sec:genmethod}, we discuss briefly the field equations of
supergravity.

Since a natural question to ask is how much supersymmetry is preserved by
any brane solution, we divide our solutions into those that preserve some of
the supersymmetry (section \ref{sec:susy}) and those that do not (section %
\ref{sec:nonsusy}). We begin by considering some four-dimensional metrics
with self-dual curvature. We review the embedding of the Taub-NUT solution
in four dimensions (TN$_{4}$), including its supersymmetric properties. We
obtain a solution dual to that obtained for the D2/D6 intersecting brane
system found previously \cite{Hashi}. Another metric with similar self-dual
properties is the Eguchi-Hanson (EH) metric, which we consider in section %
\ref{sec:susyeh4}; we obtain two new D2/D6 intersecting brane solutions for
this case. We then discuss embedding products of these metrics: TN$%
_{4}\otimes $TN$_{4}$ (section \ref{sec:susytn4xtn4}), EH$\otimes $EH
(section \ref{sec:susyeh4xeh4}) and TN$_{4}\otimes $EH (section \ref%
{sec:susytn4xeh4}). All of these solutions preserve some of the
supersymmetry, and for completeness we discuss this in an appendix.

We then consider embedding variants of the Taub-NUT spaces. Here we find a
rich variety of solutions, though they do\ not preserve any supersymmetry.
In section \ref{sec:4dtb}, we consider the embedding of four-dimensional
Taub-Bolt space in M-theory. In this case, we show that the partial
differential equation can again be separated into two ordinary differential
equations. The solution of one of these differs from the Taub-NUT case \cite%
{Hashi}; although an analytic solution is not possible, we find a numerical
solution of the other equation. In section \ref{sec:6dtntb}, we consider the
embedding of six-dimensional Taub-NUT and Taub-Bolt spaces with one NUT
charge. Again we find that the partial differential equations can be
separated into two ordinary differential equations: one can be solved
analytically, and the other one has a proper solution, which we find through
numerical methods. In all the above mentioned cases, we find the different
type IIA supergravity NSNS and RR fields. The ten-dimensional metric
obtained from reduction of the four-dimensional Taub-Bolt is a D2/D6 system.
In the case of six-dimensional Taub-NUT(Bolt), we find a D2 brane localized
in the world-volume of a D4 brane. This solution is a new intersecting brane
solution, different from the known intersecting brane obtained by
T-dualizing the D-brane solutions \cite{behr,gau}. In section \ref%
{sec:8dtntb}, we consider the embedding of the eight-dimensional Taub-NUT
and Taub-Bolt spaces with one NUT charge. In these cases, we consider an
M-brane solution in a transverse background with one non-compact transverse
direction and find exact analytic solutions.

\section{General Method}

\label{sec:genmethod}

We begin with the general Lagrangian from which the equations of
11-dimensional supergravity can be constructed \cite{DuffKK}, briefly
reviewing the equations we need here (and employing the same notation). In
order to construct a solution to these equations that can be successfully
reduced to $D=10$ dimensional type IIA string theory, we assume a bosonic
ground state, i.e. the VEV of any fermion field should be zero. This will
allow us to focus on the equations for $g_{MN}$ and $A_{MNP}$, which are now
given by 
\begin{eqnarray}
R_{MN}-\frac{1}{2}g_{MN}R &=&\frac{1}{3}\left[ F_{MPQR}F_{N}^{~PQR}-\frac{1}{%
8}g_{MN}F_{PQRS}F^{PQRS}\right]   \label{GminG2} \\
\nabla _{M}F^{MNPQ} &=&-\frac{1}{576}\varepsilon ^{M_{1}\ldots
M_{8}NPQ}F_{M_{1}\ldots M_{4}}F_{M_{5}\ldots M_{8}}  \label{dFgen}
\end{eqnarray}%
where because $\left\langle \Psi _{M}\right\rangle =0$, $F_{MNPQ}$ is the
unmodified four-form field strength 
\begin{eqnarray}
F_{MNPQ} &=&4\partial _{\lbrack M}A_{NPQ]}  \notag \\
&=&\frac{1}{2}\left[ A_{MNP,Q}-A_{NPQ,M}+A_{PQM,N}-A_{QMN,P}\right] .
\label{Fgen}
\end{eqnarray}%
Note that in (\ref{dFgen}) the $\varepsilon $ tensor is an 11-component,
11-dimensional tensor, whose usage involves memory-intensive computer
algebra calculations. In our case, equation (\ref{dFgen}) is easily shown to
be satisfied by all our solutions: due to the non-zero components of the
three-form $A_{MNP}$ (or equivalently, non-zero components of the four-form $%
F_{MNPQ}$; see eq. (\ref{Ft12yr}) for more details), the right-hand side of (%
\ref{dFgen}) is zero, so we need only show that the left-hand side also
vanishes.

The general ansatz for an M2 brane solution \cite{gauntlett} takes the form 
\begin{equation}
ds^{2}=H^{1/3}\left[ H^{-1}(-dt^{2}+dx_{1}^{2}+dx_{2}^{2})+\left( ds_{\text{%
mtrc}1}^{2}+ds_{\text{mtrc}2}^{2}\right) \right]  \label{ansatz11d}
\end{equation}%
where in general $H$ depends on the coordinates transverse to the brane $%
H=H(x_{3},\ldots ,x_{10})$, and $A_{tx_{1}x_{2}}=1/H$. We have labelled the
eight-dimensional space that is not part of the brane world-volume in (\ref%
{ansatz11d}) by two metrics $ds_{\text{mtrc}1}^{2}$ and $ds_{\text{mtrc}%
2}^{2}$. We shall take these to be any combination of the following forms:
flat space, $k$-dimensional Taub-NUT/Bolt, or Eguchi-Hanson. By embedding a $%
k$ dimensional (Euclideanized) self-dual metric (i.e. a Taub-NUT or
Eguchi-Hanson metric) into this equation, we already achieve the form
required for Kaluza-Klein reduction of one of the coordinates on a circle.
We can then calculate the left-hand side of (\ref{GminG2}) for specific
forms of this metric, and using (\ref{Fgen}), calculate the right-hand side
and thereby compute $H$. As in ref. \cite{Hashi}, we will only take $H$ to
depend on at most two of the transverse coordinates.

From (\ref{ansatz11d}), one can use 
\begin{equation}
g_{AB}=\left[ 
\begin{array}{cc}
e^{-2\Phi /3}\left( g_{\alpha \beta }+e^{2\Phi }C_{\alpha }C_{\beta }\right)
& \nu e^{4\Phi /3}C_{\alpha } \\ 
\nu e^{4\Phi /3}C_{\beta } & \nu ^{2}e^{4\Phi /3}%
\end{array}%
\right]  \label{genkkmetric}
\end{equation}%
where $\nu $ is the winding number, giving the number of times the membrane
wraps around the compactified dimension \cite{Townsend}. For simplicity we
will take $\nu =1$\ in what follows. From (\ref{genkkmetric}) and the
reduction of $A_{MNP}$ to its ten dimensional form, the Ramond-Ramond (RR) ($%
C_{\alpha }$, $A_{\alpha \beta \gamma }$) and Neveu-Schwarz Neveu-Schwarz
(NSNS) ($\Phi $, $B_{\alpha \beta }$ and $g_{\alpha \beta }$) fields can be
easily read off. Once the ten dimensional equations are found, their
analysis and comparison to existing forms can be carried out. \ In obtaining
the relation (\ref{genkkmetric}) with $\nu =1$, we use the well known
Kaluza-Klein reduction of the 11D supergravity metric and field strength to
10D metric and field strength \cite{smith} 
\begin{eqnarray}
ds_{(1,10)}^{2} &=&e^{-2\Phi /3}ds_{(1,9)}^{2}+e^{4\Phi
/3}(dx_{10}+C_{\alpha }dx^{\alpha })^{2}  \label{KKred} \\
F_{(4)} &=&\mathcal{F}_{(4)}+H_{(3)}\wedge dx_{10}  \label{KKred2}
\end{eqnarray}%
where $\mathcal{F}_{(4)}$ and $H_{(3)}$\ are the RR four-form and NSNS
three-form field strengths corresponding to $A_{\alpha \beta \gamma }$ and $%
B_{\alpha \beta }$ and $x_{10}$\ is the coordinate of compactified manifold.
We take it to be a circle with radius $R_{\infty }$, parameterized as $%
x_{10}=R_{\infty }\psi $\ where $\psi $\ has period $2\pi .$\ Although we
have assumed $\nu =1,$ \ the $\nu \neq 1$ case can be dealt with by
compactifying $\nu $\ times over this circle and replacing $x_{10}$\ by $\nu
x_{10}$\ in the relations (\ref{KKred}) and (\ref{KKred2}). This simply adds
to the dilaton field a constant term of the form $\frac{3}{2}\ln \nu ,$\ and
multiplies the RR field $C_{\alpha }$ by a multiplicative constant of $\frac{%
1}{\nu }$ when we reduce the theory to 10 dimensions.{\large \ }

Preservation of supersymmetry is checked by finding non-trivial solutions to
the Killing spinor equation \cite{gauntlett} 
\begin{equation}
\partial _{M}\epsilon +\frac{1}{4}\omega _{Mab}\Gamma ^{ab}\epsilon +\frac{1%
}{144}\Gamma _{M}^{\phantom{M}NPQR}F_{NPQR}\epsilon -\frac{1}{18}\Gamma
^{PQR}F_{MPQR}\epsilon =0  \label{genkspinor}
\end{equation}%
where $\epsilon $\ is the anticommuting parameter of the supersymmetry
transformation and the indices $M,N,\ldots $ are $d=11$ world indices, while 
$a,b,\ldots $ are $d=11$ tangent space indices. Our conventions are 
\begin{eqnarray}
de^{a} &=&g_{\phantom{a}bc}^{a}e^{b}\wedge e^{c} \\
\omega _{\phantom{a}bc}^{a} &=&\frac{1}{2}\left( g_{\phantom{a}bc}^{a}+g_{%
\phantom{b}ca}^{b}-g_{\phantom{c}ab}^{c}\right) \\
\omega _{dbM} &=&\omega _{\phantom{a}bc}^{a}\eta _{ad}e_{\phantom{d}M}^{c}
\end{eqnarray}%
where the usual definitions and properties 
\begin{equation}
e^{a}=e_{\phantom{a}M}^{a}dx^{M}\text{ \ \ \ \ \ \ \ \ }g_{MN}=\eta
_{ab}e^{a}e^{b}
\end{equation}%
(etc.) hold. The $\Gamma $'s in (\ref{genkspinor}) are given by 
\begin{equation}
\Gamma ^{A_{1}\ldots A_{p}}=\Gamma ^{\lbrack A_{1}}\ldots \Gamma ^{A_{p}]}
\label{Gamma1top}
\end{equation}%
and in general we denote $\Gamma ^{ab}=\Gamma ^{\lbrack a}\Gamma ^{b]}$. The
gamma matrices must also satisfy the Clifford Algebra 
\begin{equation}
\left\{ \Gamma ^{a},\Gamma ^{b}\right\} =-2\eta ^{ab}  \label{CliffordAlg}
\end{equation}%
where we are using the Lorentzian signature $\left[ -1,+1,\ldots ,+1\right] $%
. A representation of the algebra (\ref{CliffordAlg}) is given by 
\begin{equation}
\Gamma _{\widehat{i}}=\gamma _{\widehat{i}}\otimes 1_{8}  \label{GAmmaihat}
\end{equation}%
and 
\begin{equation}
\Gamma _{\xi +4}=\gamma _{5}\otimes \widehat{\Gamma }_{\xi }
\end{equation}%
where $\widehat{i}=0,1,2,3$ and $\xi =0,1,...,6$\ denotes the spacetime
indices for the tangent space groups $SO(1,3)$\ and $SO(7)$. The $\Gamma
_{\xi +4}$\ (and $\widehat{\Gamma }_{\xi }$) satisfy the anticommutation
relations 
\begin{equation}
\{\Gamma _{\xi +4}^{\text{ \ \ }},\Gamma _{\zeta +4}^{\text{ \ \ }}\}=\{%
\widehat{\Gamma }_{\xi }^{\text{ \ \ }},\widehat{\Gamma }_{\zeta }^{\text{ \
\ }}\}=-2\delta _{\xi \zeta }  \label{Cliff8}
\end{equation}%
where the $\widehat{\Gamma }_{\xi }$'s are given by{\large \ } 
\begin{equation}
\begin{array}{c}
\widehat{\Gamma }_{0}=i\gamma _{0}\otimes 1_{2} \\ 
\widehat{\Gamma }_{i}=\gamma _{i}\otimes 1_{2} \\ 
\widehat{\Gamma }_{i+3}=i\gamma _{5}\otimes \sigma _{i}%
\end{array}
\label{Gammahats}
\end{equation}%
in terms of the Pauli matrices $\sigma _{i}$ $(i=1,2,3)$, $\gamma
_{0}=\left( 
\begin{array}{cc}
0 & 1_{16} \\ 
1_{16} & 0%
\end{array}%
\right) $, and $\gamma _{5}=i\gamma _{0}\gamma _{1}\gamma _{2}\gamma _{3}$.

\section{Supersymmetric solutions}

\label{sec:susy}

\subsection{Four Dimensional Taub-NUT solution Revisited}

\label{sec:susytn4}

We review here the embedding of the four-dimensional Taub-NUT solution \cite%
{Hashi}. The eleven dimensional metric is given by 
\begin{eqnarray}
ds_{11}^{2} &=&H(y,r)^{-2/3}\left( -dt^{2}+dx_{1}^{2}+dx_{2}^{2}\right)
+H(y,r)^{1/3}\left( dy^{2}+y^{2}d\alpha _{1}^{2}\right. +  \notag \\
&&+\left. y^{2}\sin ^{2}(\alpha _{1})d\alpha _{2}^{2}+y^{2}\sin ^{2}(\alpha
_{1})\sin ^{2}(\alpha _{2})d\alpha _{3}^{2}+ds_{TN_{4}}^{2}\right) 
\label{ds11tn4} \\
F_{tx_{1}x_{2}y} &=&-\frac{1}{2H^{2}}\frac{\partial H}{\partial y}%
~~,~~~F_{tx_{1}x_{2}r}=-\frac{1}{2H^{2}}\frac{\partial H}{\partial r}
\label{Ft12yr}
\end{eqnarray}%
where $ds_{TN_{4}}^{2}$ can be written in either the form given by \cite%
{Hashi} 
\begin{eqnarray}
ds_{TN_{4}}^{2} &=&\tilde{f}_{4}(r)\left( dr^{2}+r^{2}(d\theta ^{2}+\sin
^{2}(\theta )d\phi ^{2})\right) +\left( \frac{(4n)^{2}}{\tilde{f}_{4}(r)}%
\right) \left( d\psi +\frac{1}{2}\cos (\theta )d\phi \right) ^{2}
\label{dstn4hashi} \\
\tilde{f}_{4}(r) &=&\left( 1+\frac{2n}{r}\right) 
\end{eqnarray}%
or equivalently as 
\begin{eqnarray}
ds_{TN_{4}}^{2} &=&\frac{1}{f_{4}(r)}\left[ d\psi ^{\prime }+2n\cos (\theta
)d\phi \right] ^{2}+f_{4}(r)dr^{2}+(r^{2}-n^{2})\left( d\theta ^{2}+\sin
^{2}(\theta )d\phi ^{2}\right) ~~~  \label{dstn4} \\
f_{4}(r) &=&\frac{(r+n)}{(r-n)}.  \label{f4tn}
\end{eqnarray}%
One can go from (\ref{dstn4}) to (\ref{dstn4hashi}) by letting $r\rightarrow
r+n$ and $\psi ^{\prime }=4n\psi $.

Requiring that (\ref{ds11tn4}) and (\ref{Ft12yr}) satisfy the field
equations entails calculating both sides of (\ref{GminG2}) and then solving
the resulting differential equation 
\begin{equation}
\frac{3(r+2n)}{y}\frac{\partial H(y,r)}{\partial y}+(r+2n)\frac{\partial
^{2}H(y,r)}{\partial y^{2}}+2\frac{\partial H(y,r)}{\partial r}+r\frac{%
\partial ^{2}H(y,r)}{\partial r^{2}}=0  \label{diffeqtn4}
\end{equation}%
for $H(y,r)$. By substituting 
\begin{equation}
H(y,r)=1+Q_{M2}Y(y)R(r)  \label{Hyrsoln}
\end{equation}%
where $Q_{M2}$ is the M2 brane charge, we obtain 
\begin{equation}
\frac{\partial ^{2}Y}{\partial y^{2}}+\frac{3}{y}\frac{\partial Y}{\partial y%
}+p^{2}Y=0  \label{BesselJeqn}
\end{equation}%
whose solution is $J_{1}(py)/y$. This yields 
\begin{equation}
H_{TN_{4}}(y,r)=1+Q_{M2}\int_{0}^{\infty }dp\frac{(py)^{2}J_{1}(py)}{4\pi
^{2}y^{3}}R_{p}(r)  \label{hashiHyr}
\end{equation}%
where 
\begin{equation}
R_{p}(r)=\frac{\pi ^{2}p}{16}\Gamma (pn)\frac{\mathcal{W}_{W}(-pn,1/2,2pr)}{r%
}=\frac{\pi ^{2}p^{2}}{8}\Gamma (pn)e^{-pr}\mathcal{U}(1+pn,2,2pr)
\label{RTN4}
\end{equation}%
is the solution to 
\begin{equation}
r\frac{d^{2}R_{p}(r)}{dr^{2}}+2\frac{dR_{p}(r)}{dr}-p^{2}\left( r+2n\right)
R_{p}(r)=0.  \label{RDETN4}
\end{equation}

Here $\mathcal{W}_{W}(x,y,z)$\ is the Whittaker-Watson function, related to $%
\mathcal{U}$, the Kummer U-function or confluent hypergeometric function, as
noted above. Reducing (\ref{ds11tn4}) down to ten dimensions gives rise to
the type IIA supergravity solution of a D2 brane localized along a D6 brane
mentioned in the introduction \cite{Hashi}.

\begin{figure}[tbp]
\centering                       
\begin{minipage}[c]{.3\textwidth}
        \centering
        \includegraphics[width=\textwidth]{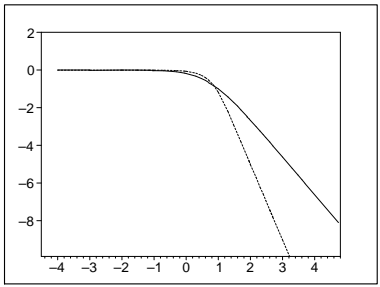}
    \end{minipage}
\caption{Log-Log plots of the functions $h(r)$ (solid)$\ $and $\widetilde{h}%
(y)$ (dotted) in terms of $\frac{r}{n}$ and $\frac{y}{n},$ respectively.}
\label{figHTN4s}
\end{figure}

In addition to the solution (\ref{hashiHyr}), we easily obtain another
solution by changing the separation constant $p\rightarrow ic$. In this case
the radial equation becomes 
\begin{equation}
r\frac{d^{2}R_{c}(r)}{dr^{2}}+2\frac{dR_{c}(r)}{dr}+c^{2}\left( r+2n\right)
R_{c}(r)=0  \label{DETN4sc}
\end{equation}%
and its solution is given by

\begin{equation}
R_{c}(r)=C_{c}\frac{(-i)\mathcal{W}_{M}(-icn,1/2,2icr)}{r}=D_{c}e^{-icr}%
\text{\ }\mathcal{G}(1+icn,2,2icr)  \label{SolDETN4sc}
\end{equation}%
where $\mathcal{G}$\ is a hypergeometric function which is finite at $r=0$,
and undergoes damped oscillations until it vanishes at $r=\infty $. Here $%
C_{c}$ or $D_{c}$ are constants that depend only on $n$ and the separation
constant $c$. The solution of the modified differential equation (\ref%
{BesselJeqn}) for $Y(y)$ is now

\begin{equation}
Y_{c}(y)=E_{c}\frac{K_{1}(cy)}{y}  \label{SWOlDE2TN4sc}
\end{equation}%
where $K_{1}$\ is the modified Bessel function, diverging at $y=0$ and
vanishing at infinity. Writing the general solution of the metric function
as a superposition of the solutions with constant $c$, we have 
\begin{equation}
\widetilde{H}_{TN_{4}}(y,r)=1+Q_{M2}\int_{0}^{\infty
}dcR_{c}(r)Y_{c}(y)=1+Q_{M2}\int_{0}^{\infty }dcf(c)e^{-icr}\mathcal{G}%
(1+icn,2,2icr)\frac{K_{1}(cy)}{y}  \label{HM2secondcase}
\end{equation}%
where $f(c)=D_{c}E_{c}$ and we choose $c$ to belong to the interval $%
[0,\infty )$. The function $f(c)$ may be determined from a consideration of
the near horizon limit. In this limit, where $r<<n$ (or equivalently when $%
n\rightarrow \infty $\ and $r\rightarrow 0$), the metric (\ref{dstn4})
reduces to $\mathbb{R}^{4}$, with the line element 
\begin{equation}
ds^{2}=dz^{2}+z^{2}d\Omega _{3}^{2}  \label{TN4lim}
\end{equation}%
where $z=2\sqrt{2nr}.$\ The transverse geometry is then $\mathbb{R}%
^{4}\otimes \mathbb{R}^{4}$ with the metric 
\begin{equation}
ds^{2}=dy^{2}+dz^{2}+(y^{2}+z^{2})d\Omega _{3}^{2}  \label{transflat}
\end{equation}%
and the metric function (\ref{HM2secondcase}) should coincide with $1+\frac{%
Q }{R^{6}}$, where 
\begin{equation}
R=\sqrt{y^{2}+z^{2}}  \label{R}
\end{equation}%
Hence we get%
\begin{eqnarray}
&&\lim_{z^{2}<<8n^{2}}\int_{0}^{\infty }dcf(c)e^{-ic\frac{z^{2}}{8n}}%
\mathcal{G}(1+icn,2,ic\frac{z^{2}}{4n})\frac{K_{1}(cy)}{y}  \notag \\
&=&\int_{0}^{\infty }dcf(c)\frac{I_{1}(2ic\sqrt{2nr})}{ic\sqrt{2nr}}\frac{%
K_{1}(cy)}{y}=\frac{1}{(y^{2}+8nr)^{3}}  \label{HTN4lim}
\end{eqnarray}%
{\large \bigskip }which yields $f(c)=\frac{c^{4}}{16}$, or

\begin{equation}
\widetilde{H}_{TN_{4}}(y,r)=1+\frac{Q_{M2}}{16}\int_{0}^{\infty
}dcc^{4}e^{-icr}\mathcal{G}(1+icn,2,2icr)\frac{K_{1}(cy)}{y}.
\label{TN4gfun}
\end{equation}%
This same approach can be used to determine the form of the integral in
equation (\ref{hashiHyr}), though we note that a different method was
employed in ref. \cite{Hashi}. In figure \ref{figHTN4s}, the log-log plots
of $h(r)\simeq (H_{TN_{4}}(y=0,r)-1)$ versus $\frac{r}{n}$\ and $\widetilde{h%
}(y)\simeq (\widetilde{H}_{TN_{4}}(y,r=0)-1)$ versus $\frac{y}{n}$, are
given where we choose the normalization coefficients such that two functions
approach one as $r$ and $y$ go to zero. The two functions have the same
behaviour qualitatively, though $\widetilde{h}(y)$ has a considerably
steeper falloff. 

By inserting either of the solutions (\ref{hashiHyr}) or (\ref{TN4gfun})
into the metric (\ref{ds11tn4}), it is straightforward to show that both
solutions preserve 1/4 of the supersymmetry, as we demonstrate in appendix %
\ref{sec:susycalc}.

Dimensional reduction of (\ref{ds11tn4}) with (\ref{hashiHyr}) or (\ref%
{TN4gfun}) along the coordinate $\psi $\ of the Taub-NUT$_{4}$\ metric gives
the type IIA supergravity metric%
\begin{eqnarray}
ds_{10}^{2} &=&H^{-1/2}\widetilde{f}_{4}^{-1/2}\left(
-dt^{2}+dx_{1}^{2}+dx_{2}^{2}\right) +  \notag \\
&+&H^{1/2}\widetilde{f}_{4}^{-1/2}\left( dy^{2}+y^{2}d\Omega _{3}^{2}\right)
+H^{1/2}\widetilde{f}_{4}^{1/2}(dr^{2}+r^{2}d\Omega _{2}^{2})
\label{ds10TN4}
\end{eqnarray}%
which describes a D2 brane localized along the world-volume of D6 brane. The
other fields in ten dimensions are NSNS fields 
\begin{eqnarray}
\Phi  &=&\frac{3}{4}\ln \left\{ \frac{H^{1/3}}{\widetilde{f}_{4}}\right\}  \\
B_{\mu \nu } &=&0
\end{eqnarray}%
and Ramond-Ramond (RR) fields 
\begin{eqnarray}
C_{\phi } &=&2n\cos (\theta ) \\
A_{tx_{1}x_{2}} &=&\frac{1}{H}
\end{eqnarray}%
where in the above relations, $H$ refers either to $H_{TN_{4}}(y,r)$ or $%
\widetilde{H}_{TN_{4}}(y,r)$ given by (\ref{hashiHyr}) and (\ref{TN4gfun}).

\subsection{4D Eguchi-Hanson}

\label{sec:susyeh4}

The Eguchi-Hanson metric is another asymptotically flat metric with
self-dual curvature that can also be embedded into eleven dimensions,
furnishing a localized brane solution similar to the Taub-NUT metric above.
The eleven dimensional metric will be the same as (\ref{ds11tn4}), but with $%
ds_{TN_{4}}^{2}$ replaced with \cite{Mahapatra} 
\begin{eqnarray}
ds_{EH}^{2} &=&\frac{r^{2}}{4g(r)}\left[ d\psi +\cos (\theta )d\phi \right]
^{2}+g(r)dr^{2}+\frac{r^{2}}{4}\left( d\theta ^{2}+\sin ^{2}(\theta )d\phi
^{2}\right)   \label{dseh4} \\
g(r) &=&\left( 1-\frac{a^{4}}{r^{4}}\right) ^{-1}.
\end{eqnarray}%
The four-form field strength is given again by (\ref{Ft12yr}). We find that
this new eleven dimensional metric will satisfy the supergravity field
equations, provided 
\begin{equation}
\frac{(3r^{4}+a^{4})}{r^{5}}\frac{\partial H}{\partial r}+\frac{%
(r-a)(r+a)(r^{2}+a^{2})}{r^{4}}\frac{\partial ^{2}H}{\partial r^{2}}+\frac{3%
}{y}\frac{\partial H}{\partial y}+\frac{\partial ^{2}H}{\partial y^{2}}=0.
\label{eh4diffeq1}
\end{equation}%
This equation is separable, and becomes 
\begin{eqnarray}
\frac{(r^{4}-a^{4})}{r^{4}}\frac{d^{2}R_{c}(r)}{dr^{2}}+\frac{(3r^{4}+a^{4})%
}{r^{5}}\frac{dR_{c}(r)}{dr}-c^{2}R_{c}(r) &=&0  \label{eh4diffeq2R} \\
\frac{d^{2}Y(y)}{dy^{2}}+\frac{3}{y}\frac{dY(y)}{dy}+c^{2}Y(y) &=&0
\label{eh4diffeq2Y}
\end{eqnarray}%
upon substituting in $H(y,r)=1+Q_{M2}Y(y)R_{c}(r)$, where $c^{2}$ is a
separation constant. Eq. (\ref{eh4diffeq2Y}) is identical to equation (\ref%
{BesselJeqn}) and so we again obtain 
\begin{equation}
Y(y)\sim \frac{J_{1}(cy)}{y}.  \label{eh4Y}
\end{equation}

\begin{figure}[tbp]
\centering                       
\begin{minipage}[c]{.3\textwidth}
        \centering
        \includegraphics[width=\textwidth]{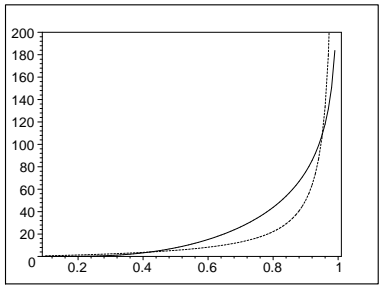}
    \end{minipage}
\caption{Numerical solutions $R_{c}/10^{48}$ of the radial equation (\ref%
{eh4diffeq2R}) as a function of $\frac{a}{r}$ (solid)$,$ and solution $%
R_{p}/10^{6}$ of the radial equation (\ref{RDETN4}) for TN$_{4}$ as a
function of $\frac{n}{r}$ (dotted). For $r\approx a$ (or $n$), both $R$'s
diverge and for $r\approx \infty $, they vanish.}
\label{fig0}
\end{figure}

We can solve the equation for $R_{c}(r)$ numerically. For large $r$, the
solution of the equation (\ref{eh4diffeq2R}) that vanishes at infinity is $%
\frac{K_{1}(cr)}{r}$, where $K_{1}$ is the modified Bessel function. A
typical numerical solution to the radial differential equation (\ref%
{eh4diffeq2R}) versus $a/r$ is displayed in figure \ref{fig0}. Note that in
the metric (\ref{dseh4}), the coordinate $r$ must be greater than or equal
to $a$. For comparison, a plot of the $R_{p}(r)$ given by the relation (\ref%
{RTN4}) is also plotted in figure \ref{fig0}; qualitatively they are the
same, though the EH function has a less rapid falloff. Roughly speaking, the
Eguchi-Hanson parameter $a$ plays the same role of the NUT charge $n$ in the
TN$_{4}$ case. 

We obtain 
\begin{equation}
H(y,r)=1+Q_{M2}\int_{0}^{\infty }dcg(c)\frac{J_{1}(cy)}{y}R_{c}(r).
\label{HEH4}
\end{equation}%
To fix the measure function $g(c)$, we compare the above relation to that of
a metric function of a brane in an 8-dimensional flat metric\ $\mathbb{R}%
^{4}\otimes \mathbb{R}^{2}\otimes S^{2},$ obtained by looking at the near
horizon limit. We note that for $r=a(1+\epsilon ^{2})$, where $\epsilon <<1,$%
\ the metric reduces to 
\begin{eqnarray}
ds_{r=a(1+\epsilon ^{2})}^{2} &=&a^{2}\{\epsilon ^{2}\left[ d\psi +\cos
(\theta )d\phi \right] ^{2}+d\epsilon ^{2}\}+\frac{a^{2}}{4}\left( d\theta
^{2}+\sin ^{2}(\theta )d\phi ^{2}\right)   \label{dseh4reqa} \\
&\approx &z^{2}d\psi ^{2}+dz^{2}+\frac{a^{2}}{4}\left( d\theta ^{2}+\sin
^{2}(\theta )d\phi ^{2}\right) 
\end{eqnarray}%
which is $\mathbb{R}^{2}\otimes S^{2}$\ with the radial length equal to $%
\sqrt{z^{2}+\frac{a^{2}}{4}}$. If we assume that the parameter $a$ is small,
the differential equation (\ref{eh4diffeq2R}) reduces to 
\begin{equation}
\ddot{R}_{c}+\frac{3}{\widehat{r}}\dot{R}_{c}-c^{2}R_{c}=0
\label{eh4diffeq2Rlimit}
\end{equation}%
where $\widehat{r}=r-a=a\epsilon ^{2}\approx 0$ and the overdot denotes $%
\frac{d}{d\widehat{r}}$. This equation has the solution

\begin{equation}
R_{c}(\widehat{r})=\frac{A_{c}}{\widehat{r}}K_{1}(c\widehat{r})+\frac{B_{c}}{%
\widehat{r}}I_{1}(c\widehat{r})  \label{eh4diffeq2Rlimitsolution}
\end{equation}%
which will vanish at infinity provided $B_{c}=0$, though it will diverge at $%
\widehat{r}=0$.

Taking these limits into account in equation (\ref{HEH4}), we find 
\begin{eqnarray}
\int_{0}^{\infty }dcg(c)\frac{K_{1}(c\widehat{r})}{\widehat{r}}J_{1}(cy)
&=&\lim_{a\rightarrow 0}\frac{y}{(z^{2}+\frac{a^{2}}{4}+y^{2})^{3}}  \notag
\\
&=&\lim_{a\rightarrow 0}\frac{y}{(\widehat{r}a+y^{2})^{3}}=\frac{1}{y^{5}}
\label{HEHlim}
\end{eqnarray}%
where we can absorb the constant $A_{c}$ into the definition of $g(c)$. By
comparing the above relation with the known integral 
\begin{equation}
\int_{0}^{\infty }dcc^{3}\frac{K_{1}(c\widehat{r})}{\widehat{r}}J_{1}(cy)=%
\frac{8y}{(\widehat{r}^{2}+y^{2})^{3}}\overset{\widehat{r}=a\epsilon ^{2}}{%
\rightarrow }\frac{8}{y^{5}}  \label{K1J1int}
\end{equation}%
we find that $g(c)=\frac{c^{4}}{8}$, and so the metric function becomes 
\begin{equation}
H_{EH}(y,r)=1+Q_{M2}\int_{0}^{\infty }dc\left( \frac{c^{4}}{8}\right)
R_{c}(r)\frac{J_{1}(cy)}{y}.  \label{HfinalEH4}
\end{equation}

\begin{figure}[tbp]
\centering                       
\begin{minipage}[c]{.3\textwidth}
        \centering
        \includegraphics[width=\textwidth]{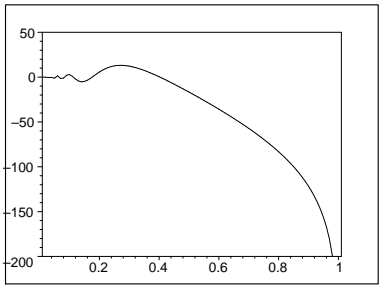}
    \end{minipage}
\caption{{} Numerical solution of radial equation (\ref{eh4xeh4deqR2}) for $%
\mathcal{R}_{2}/10^{5}$ (or $\widetilde{R}_{c}/10^{5}$) as a function of $%
\frac{1}{r_{2}}$ (or $\frac{1}{r}$) for non-zero separation constant. The
Eguchi-Hanson parameters $a_{2}$ ($a$) are set to one and so for $%
r_{2}\approx a_{2}$ ($r\approx a$)$,$ the function $\mathcal{R}_{2}$ ($%
\widetilde{R}_{c}$) diverges and for $r_{2}\approx \infty $ ($r\approx
\infty $), it vanishes.}
\label{fig011}
\end{figure}

By changing $c\rightarrow ic$ in the differential equations (\ref%
{eh4diffeq2R}) and (\ref{eh4diffeq2Y}), we get another solution in the form
of 
\begin{equation}
\widetilde{H}_{EH}(y,r)=1+Q_{M2}\int_{0}^{\infty }dc\left( \frac{c^{4}}{8}%
\right) \widetilde{R}_{c}(r)\frac{K_{1}(cy)}{y}  \label{HEH4sc}
\end{equation}%
where $\widetilde{R}_{c}(r)$\ is a damped oscillating function whose
behaviour is plotted in figure \ref{fig011} and $\widetilde{g}(c)=\frac{c^{4}%
}{8}$. 

Equations (\ref{HfinalEH4}) and (\ref{HEH4sc}) are the respective analogues
of the TN$_{4}$ solutions (\ref{hashiHyr}) and (\ref{TN4gfun}) for the EH
case. They can be plotted numerically to yield results qualitatively similar
to those obtained for $H_{TN4}$ and $\widetilde{H}_{TN4}$ in the previous
section; we will not present them here.

Reduction of these solutions to a ten dimensional type IIA string theory
solution proceeds in a manner similar to the reduction of $S^{4}$ regarded
as containing a NUT and anti-NUT charge \cite{ChamGib}. \ The vector $%
\partial /\partial \psi $ generates a Hopf fibration of a 3-sphere in the
metric (\ref{dseh4}). Using (\ref{genkkmetric}) we obtain the NSNS fields%
\begin{equation}
\begin{array}{rcl}
\Phi  & = & \frac{3}{4}\ln \left\{ \frac{H^{1/3}w^{2}}{4g}\right\}  \\ 
B_{\mu \nu } & = & 0%
\end{array}
\label{NSNSEH}
\end{equation}%
where we define the dimensionless coordinate $w$ by $w=\frac{r}{a}$. The
Ramond-Ramond (RR) fields and the ten dimensional metric will be given by 
\begin{equation}
\begin{array}{rcl}
C_{\phi } & = & a\cos (\theta ) \\ 
A_{tx_{1}x_{2}} & = & \frac{1}{H}%
\end{array}
\label{RREH}
\end{equation}%
\begin{eqnarray}
ds_{10}^{2} &=&\frac{w}{2}\{H^{-1/2}g^{-1/2}\left(
-dt^{2}+dx_{1}^{2}+dx_{2}^{2}\right) +  \notag \\
&+&H^{1/2}g^{-1/2}\left( dy^{2}+y^{2}d\Omega _{3}^{2}\right)
+H^{1/2}g^{1/2}a^{2}(dw^{2}+\frac{w^{2}}{4g}d\Omega _{2}^{2})\}.
\label{dsEH10}
\end{eqnarray}%
The metric (\ref{dsEH10}) describes a D2/D6 system where the D2-brane is
localized along the world-volume of the D6-brane. We note $H$ refers to
either $H_{EH}$ or $\widetilde{H}_{EH}$ and $g=g(w)=\left( 1-w^{-4}\right)
^{-1}$ in the above relations. We have explicitly checked that the above
10-dimensional metric, with the given dilaton and one form, is a solution to
the 10-dimensional supergravity equations of motion.

The metric (\ref{dsEH10}) is locally asymptotically flat (though the dilaton
field diverges); for large $w$ it\ reduces to 
\begin{equation}
ds_{10}^{2}=\frac{w}{2}\{-dt^{2}+dx_{1}^{2}+dx_{2}^{2}+dy^{2}+y^{2}d\Omega
_{3}^{2}+a^{2}(dw^{2}+\frac{w^{2}}{4}d\Omega _{2}^{2})\}
\label{ds10Ehlarger}
\end{equation}%
which is a 10D locally flat metric with solid deficit angles. The Kretchmann
invariant of this spacetime vanishes at infinity and is given by 
\begin{equation}
R_{\mu \nu \rho \sigma }R^{\mu \nu \rho \sigma }=\frac{224}{a^{4}w^{6}}
\label{KEHlarger}
\end{equation}%
and all the components of the Riemann tensor in the orthonormal basis have
similar $\frac{1}{w^{3}}$ behaviour, vanishing at infinity.

To calculate how much supersymmetry is preserved by this solution in eleven
dimensions, we again use the spinor equation (\ref{genkspinor}). Half of the
supersymmetry is again removed via the projection operator $(1+\Gamma ^{\hat{%
t}\hat{x}_{1}\hat{x}_{2}})\epsilon =0$, due to the presence of the brane,
and another half is removed due to the self-dual nature of the EH metric, as
shown in appendix \ref{sec:susycalc}. Hence embedding the EH metric into an
eleven dimensional M2 brane metric preserves 1/4 of the supersymmetry (see
appendix \ref{sec:susycalc} for details).

{\large \bigskip }

\subsection{Taub-NUT$_{4}$ $\otimes $ Taub-NUT$_{4}$}

\label{sec:susytn4xtn4}

We can also embed two four dimensional metrics into the eleven dimensional
membrane metric. The first case we consider here is embedding two TN$_{4}$
metrics, giving the following metric: 
\begin{eqnarray}
ds_{11}^{2} &=&H(r_{1},r_{2})^{-2/3}\left(
-dt^{2}+dx_{1}^{2}+dx_{2}^{2}\right) +H(r_{1},r_{2})^{1/3}\left(
ds_{TN_{4}(1)}^{2}+ds_{TN_{4}(2)}^{2}\right)  \label{tn4xtn4mtrc} \\
ds_{TN_{4}(i)}^{2} &=&\left( \frac{(4n_{i})^{2}}{\widetilde{f}_{i}(r_{i})}%
\right) \left( d\psi _{i}+\frac{1}{2}\cos (\theta _{i})d\phi _{i}\right)
^{2}+\widetilde{f}_{i}(r_{i})\left( dr_{i}^{2}+r_{i}^{2}\left( d\theta
_{i}^{2}+\sin ^{2}(\theta _{i})d\phi _{i}^{2}\right) \right) ~~~
\label{dstn4i} \\
F_{tx_{1}x_{2}r_{i}} &=&-\frac{1}{2H^{2}}\frac{\partial H}{\partial r_{i}}
\label{tn4xtn4F012i}
\end{eqnarray}%
with $\widetilde{f}_{i}(r_{i})=\left( 1+{\textstyle\frac{2n_{i}}{r_{i}}}%
\right) $. Note that we can also use the (\ref{dstn4}) form for the Taub-NUT
metrics (which is easier for the supersymmetry check), but for reduction
down to ten dimensions the form given in equation (\ref{dstn4hashi}) is
preferable, as $\psi $ has a period of $2\pi $.

We can choose to compactify down to ten dimensions by compactifying on
either $\psi _{1}$ or $\psi _{2}$. Since either way produces the same
results (with $1\leftrightarrow 2$), we will compactify on $\psi _{2}$.
Thus, the radius $R_{\infty }$ of the circle as $r\rightarrow \infty $ with
line element $R_{\infty }\left( d\psi _{2}+{\textstyle\frac{1}{2}}\cos
\theta _{2}d\phi _{2}\right) $ is the same as in the Taub-NUT case above, 
\begin{equation}
R_{\infty }=4n_{2}.
\end{equation}%
Note that the $r_{i}$ range from $\left[ 0,\infty \right) $, the $\psi _{i}$
and the $\phi _{i}$ range over $\left[ 0,2\pi \right] $, and the $\theta _{i}
$ range over $\left[ 0,\pi \right] $.

The metric (\ref{tn4xtn4mtrc}) and (\ref{tn4xtn4F012i}) satisfies the eleven
dimensional equations of motion if the harmonic function satisfies the
differential equation 
\begin{eqnarray}
2(r_{2}+2n_{2})\frac{\partial H(r_{1},r_{2})}{\partial r_{1}}%
+r_{1}(r_{2}+2n_{2})\frac{\partial ^{2}H(r_{1},r_{2})}{\partial r_{1}^{2}}
&+&  \notag \\
+2(r_{1}+2n_{1})\frac{\partial H(r_{1},r_{2})}{\partial r_{2}}%
+r_{2}(r_{1}+2n_{1})\frac{\partial ^{2}H(r_{1},r_{2})}{\partial r_{2}^{2}}
&=&0.  \label{deqtnxtn}
\end{eqnarray}%
This equation is separable, which can again be seen by substituting in $%
H(r_{1},r_{2})=1+Q_{M2}R_{1}(r_{1})R_{2}(r_{2})$. This gives two equations
of the same form 
\begin{equation}
r_{i}\frac{\partial ^{2}R_{i}(r_{i})}{\partial r_{i}^{2}}+2\frac{\partial
R_{i}(r_{i})}{\partial r_{i}}\pm c^{2}(r_{i}+2n_{i})R_{i}(r_{i})=0
\label{tn4xtn4deqRi}
\end{equation}%
where the $\pm c^{2}$ term is present because the separable equations must
be set equal to a constant ($c^{2}$) of the opposite sign in order for (\ref%
{deqtnxtn}) to equal zero. The most general solution is the product of the
exponentially decayed and damped oscillating functions discussed in section %
\ref{sec:susytn4}, yielding%
\begin{eqnarray}
H &=&1+Q_{M2}\int_{0}^{\infty }dcR_{1c}(r_{1})R_{2c}(r_{2})  \notag \\
&=&1+Q_{M2}\int_{0}^{\infty }dch(c)e^{-icr_{1}}\mathcal{G}%
(1+icn_{1},2,2icr_{1})e^{-cr_{2}}\mathcal{U}(1+cn_{2},2,2cr_{2}).
\label{HTN4TN4}
\end{eqnarray}%
The metric of the transverse geometry in the near horizon limit reduces to $%
R^{4}\otimes R^{4}$, with the metric 
\begin{equation}
ds^{2}=dz_{1}{}^{2}+dz_{2}^{2}+z_{1}^{2}d\Omega _{3}^{2}+z_{2}^{2}d\Omega
_{3}^{^{\prime }2}  \label{8DmetrTNTNlim}
\end{equation}%
{\large \bigskip }where $z_{1}=2\sqrt{2n_{1}r_{1}},z_{2}=2\sqrt{2n_{2}r_{2}}.
$\ So we should have%
\begin{equation}
\lim_{r_{1}<<n_{1}\text{and }r_{2}<<n_{2}}\int_{0}^{\infty
}dch(c)e^{-icr_{1}}\mathcal{G}(1+icn_{1},2,2icr_{1})e^{-cr_{2}}\mathcal{U}%
(1+cn_{2},2,2cr_{2})=\frac{1}{(z_{1}^{2}+z_{2}^{2})^{3}}.  \label{HTN4TN4lim}
\end{equation}%
Using the relations for the limiting values of the hypergeometric and
confluent hypergeometric functions \cite{HandbookMathFns}, the above
relation becomes

\begin{equation}
\int_{0}^{\infty }dch(c)\frac{I_{1}(2ic\sqrt{2n_{1}r_{1}})}{c}\frac{2K_{1}(2c%
\sqrt{2n_{2}r_{2}})}{c\Gamma (cn_{2})}=\frac{i\sqrt{2n_{1}r_{1}}\sqrt{%
2n_{2}r_{2}}}{512(n_{1}r_{1}+n_{2}r_{2})^{3}}  \label{inteqforTN4TN4}
\end{equation}%
which yields{\large \ }$h(c)=\frac{c^{5}}{64}\Gamma (cn_{2})$ so that 
\begin{equation}
H=H_{\left( TN\right) ^{2}}\left( r_{1},r_{2}\right)
=1+Q_{M2}\int_{0}^{\infty }dc\frac{c^{5}}{64}\Gamma (cn_{2})e^{-icr_{1}}%
\mathcal{G}(1+icn_{1},2,2icr_{1})e^{-cr_{2}}\mathcal{U}(1+cn_{2},2,2cr_{2}).
\label{TN4TN4hfun}
\end{equation}%
Note that changing $c\rightarrow ic$\ in (\ref{tn4xtn4deqRi}) does not yield
any other solution, but instead interchanges the functions $%
R_{1}\longleftrightarrow R_{2}$. The graph of the metric function $H_{\left(
TN\right) ^{2}}\left( r_{1}=0,r_{2}\right) -1$, is exactly the same as the
graph of the function $h(r)$ in figure \ref{figHTN4s}.

Next we wish to dimensionally reduce the solution to the type IIA string
theory in ten dimensions. Doing this will give the NSNS fields 
\begin{equation}
\begin{array}{rcl}
\Phi & = & \frac{3}{4}\ln \left( \frac{H^{1/3}}{\widetilde{f}_{2}}\right) \\ 
B_{\mu \nu } & = & 0%
\end{array}
\label{tnXtnNSNS}
\end{equation}%
{\large \ }and RR fields

\begin{equation}
\begin{array}{rcl}
C_{\phi _{2}} & = & 2n_{2}\cos (\theta _{2}) \\ 
A_{tx_{1}x_{2}} & = & H^{-1}.%
\end{array}
\label{tnXtnRR}
\end{equation}%
The metric in ten dimensions will be given by 
\begin{eqnarray}
ds_{10}^{2} &=&H(r_{1},r_{2})^{-1/2}\widetilde{f}_{2}^{-1/2}\left(
-dt^{2}+dx_{1}^{2}+dx_{2}^{2}\right) +H(r_{1},r_{2})^{1/2}\widetilde{f}%
_{2}^{-1/2}\left( ds_{TN_{4}(1)}^{2}\right) +  \notag \\
&+&H(r_{1},r_{2})^{1/2}\widetilde{f}_{2}^{1/2}\left(
dr_{2}^{2}+r_{2}^{2}\left( d\theta _{2}^{2}+\sin ^{2}(\theta _{2})d\phi
_{2}^{2}\right) \right) .  \label{ds10tnXtn}
\end{eqnarray}%
This represents a D2/D6 brane system, where $H=H_{\left( TN\right) ^{2}}$ in
the preceding equation (\ref{TN4TN4hfun}).

We can further reduce the metric (\ref{ds10tnXtn}) along the $\psi _{1}$
direction of the $TN_{4}(1).${\large \ }However the result of this
compactification is not the same as the reduction of the M-theory solution (%
\ref{tn4xtn4mtrc}) over a torus, which is compactified type IIB theory. The
reason is that to get the compactified type IIB theory, we should compactify
the T-dual of the IIA metric (\ref{ds10tnXtn}) over a circle, and not
directly compactify the 10D IIA metric (\ref{ds10tnXtn}) along the $\psi
_{1} $ direction.

We note also an interesting result in reducing the 11D metric (\ref%
{tn4xtn4mtrc}) along the $\psi _{1}$ (or $\psi _{2}$) direction of the $%
TN_{4}(1)$ (or $TN_{4}(2)$) in large radial coordinates.\ As $r_{1}$ (or $%
r_{2}$) $\rightarrow \infty $\ the transverse geometry in (\ref{tn4xtn4mtrc}%
) locally approaches $\mathbb{R}^{3}\otimes S^{1}\otimes TN_{4}(2)$\ (or $%
TN_{4}(1)\otimes \mathbb{R}^{3}\otimes S^{1}$) and so the reduced theory,
obtained by compactification over the circle of the TN$_{4}$, is IIA. Then
by T-dualization of this theory (on the remaining $S^{1}$ of the transverse
geometry), we find a type IIB theory which describes the D5 defects.

This solution also preserves 1/4 of the supersymmetry. The gauge terms and
partial derivatives of $H(r_{1},r_{2})$ will remove half the supersymmetry
through the required use of (\ref{proj012}). After this, we get two sets of
equations of the form (\ref{tn4dpsi}), (\ref{tn4dtheta}) and (\ref{tn4dphi})
(with $(\psi ,\theta ,\phi )\rightarrow (\psi _{i},\theta _{i},\phi
_{i}),~i=1,2$). These can be solved by use of the equivalent of (\ref%
{projprthph}); 
\begin{eqnarray}
\Gamma ^{\hat{\psi}_{1}\hat{r}_{1}\hat{\theta}_{1}\hat{\phi}_{1}}\epsilon 
&=&\epsilon   \label{projp1r1th1ph1} \\
\Gamma ^{\hat{\psi}_{2}\hat{r}_{2}\hat{\theta}_{2}\hat{\phi}_{2}}\epsilon 
&=&\epsilon .  \label{projp2r2th2ph2}
\end{eqnarray}%
However, due to the fact that we're in eleven dimensions, (\ref{proj012})
and (\ref{projp1r1th1ph1}) imply (\ref{projp2r2th2ph2}), and so we actually
only need two projection operators, not three, meaning 1/4 of the
supersymmetry is preserved.

\subsection{Eguchi-Hanson $\otimes $ Eguchi-Hanson}

\label{sec:susyeh4xeh4}

We can also embed two Eguchi-Hanson metrics into eleven dimensions; 
\begin{eqnarray}
ds_{11}^{2} &=&H(r_{1},r_{2})^{-2/3}\left(
-dt^{2}+dx_{1}^{2}+dx_{2}^{2}\right) +H(r_{1},r_{2})^{1/3}\left(
ds_{EH1}^{2}+ds_{EH2}^{2}\right)  \\
ds_{EH(i)}^{2} &=&\frac{r_{i}^{2}}{4g_{i}(r_{i})}\left[ d\psi _{i}+\cos
(\theta _{i})d\phi _{i}\right] ^{2}+g_{i}(r_{i})dr_{i}^{2}+\frac{r_{i}^{2}}{4%
}\left( d\theta _{i}^{2}+\sin ^{2}(\theta _{i})d\phi _{i}^{2}\right) 
\label{dseh4i} \\
g_{i}(r_{i}) &=&\left( 1-\frac{a_{i}^{4}}{r_{i}^{4}}\right)
^{-1}~~~,~~~~~F_{tx_{1}x_{2}r_{i}}=-\frac{1}{2H^{2}}\frac{\partial H}{%
\partial r_{i}}.
\end{eqnarray}%
This metric will satisfy the supergravity equations if the following
differential equation holds; 
\begin{eqnarray}
\frac{(3r_{1}^{4}+a_{1}^{4})}{r_{1}^{5}}\frac{\partial H(r_{1},r_{2})}{%
\partial r_{1}}+\frac{(r_{1}-a_{1})(r_{1}+a_{1})(r_{1}^{2}+a_{1}^{2})}{%
r_{1}^{4}}\frac{\partial ^{2}H(r_{1},r_{2})}{\partial r_{1}^{2}} &+&  \notag
\\
+\frac{(a_{2}^{4}+3r_{2}^{4})}{r_{2}^{5}}\frac{\partial H(r_{1},r_{2})}{%
\partial r_{2}}+\frac{(r_{2}-a_{2})(r_{2}+a_{2})(r_{2}^{2}+a_{2}^{2})}{%
r_{2}^{4}}\frac{\partial ^{2}H(r_{1},r_{2})}{\partial r_{2}^{2}} &=&0
\end{eqnarray}%
This is again separable, using $H(r_{1},r_{2})=1+Q_{M2}\mathcal{R}_{1}(r_{1})%
\mathcal{R}_{2}(r_{2})$, and gives the following two differential equations, 
\begin{equation}
r_{1}(r_{1}^{4}-a_{1}^{4})\frac{d^{2}\mathcal{R}_{1}(r_{1})}{dr_{1}^{2}}%
+(3r_{1}^{4}+a_{1}^{4})\frac{d\mathcal{R}_{1}(r_{1})}{dr_{1}}-\mathcal{R}%
_{1}(r_{1})r_{1}^{5}c^{2}=0  \label{eh4xeh4deqR1}
\end{equation}%
\begin{equation}
r_{2}(r_{2}^{4}-a_{2}^{4})\frac{d^{2}\mathcal{R}_{2}(r_{2})}{dr_{2}^{2}}%
+(3r_{2}^{4}+a_{2}^{4})\frac{d\mathcal{R}_{2}(r_{2})}{dr_{2}}+\mathcal{R}%
_{2}(r_{2})r_{2}^{5}c^{2}=0.  \label{eh4xeh4deqR2}
\end{equation}%
These equations are the same as the $R_{c}(r)$ equation in (\ref{eh4diffeq2R}%
). Like that equation, these are not solvable analytically (unless $c=0$,
which reduces generality), but are solvable numerically. Since equation (\ref%
{eh4xeh4deqR1}) is the same as the radial differential equation (\ref%
{eh4diffeq2R}), the numerical solution of this equation given in figure \ref%
{fig0}. It vanishes at large $r_{1}$ and increases monotonically with
increasing $1/r_{1}$, diverging logarithmically at $r_{1}\approx a_{1}$.

The solution of the other differential equation (\ref{eh4xeh4deqR2}) has
damped oscillating behaviour and diverges at $r_{2}\approx a_{2}$. A typical
solution for $\mathcal{R}_{2}(r_{2}),$ is given in figure \ref{fig011}.

The most general solution is therefore given by 
\begin{equation}
H=1+Q_{M2}\int_{0}^{\infty }dcj(c)\mathcal{R}_{1c}(r_{1})\mathcal{R}%
_{2c}(r_{2}).  \label{HEH4EH4}
\end{equation}%
\bigskip The transverse geometry in the limit of small $a_{1}$\ and $a_{2}$\
\ reduces to $\mathbb{R}^{2}\otimes S^{2}\otimes \mathbb{R}^{2}\otimes S^{2}$
\ with the radial length equal to $\sqrt{z_{1}^{2}+z_{2}^{2}+\frac{a_{1}^{2}%
}{4}+\frac{a_{2}^{2}}{4}}.$\ \ In this limit, the differential equation (\ref%
{eh4xeh4deqR1}) has the solution (\ref{eh4diffeq2Rlimitsolution}) and (\ref%
{eh4xeh4deqR2}) has the solution

\begin{equation}
\mathcal{R}_{2c}(\widehat{r}_{2})=\frac{A_{2c}}{\widehat{r}_{2}}J_{1}(c%
\widehat{r}_{2})+\frac{B_{2c}}{\widehat{r}_{2}}Y_{1}(c\widehat{r}_{2})
\label{eh4diffeq2Rlimitsolution2}
\end{equation}%
which is finite at $\widehat{r}_{2}=0$\ and vanishes at infinity provided $
B_{2c}=0.$

Taking these limits into account in equation (\ref{HEH4EH4}), we must have 
\begin{eqnarray}
\int_{0}^{\infty }dc\frac{j(c)}{\widehat{r}_{1}\widehat{r}_{2}}K_{1}(c%
\widehat{r}_{1})J_{1}(c\widehat{r}_{2}) &=&\lim_{a_{1},a_{2}\rightarrow 0}%
\frac{1}{(z_{1}^{2}+z_{2}^{2}+\frac{a_{1}^{2}}{4}+\frac{a_{2}^{2}}{4})^{3}} 
\notag \\
&=&\frac{1}{(\widehat{r}_{1}a_{1}+\widehat{r}_{2}a_{2})^{3}}
\label{HEH4EH4lim}
\end{eqnarray}%
where we absorb the constants $A_{1c}$,$A_{2c}$ into the definition of $j(c)$%
. By comparing the above relation with the known integral 
\begin{equation}
\int_{0}^{\infty }dcc^{3}K_{1}(c\widehat{r}_{1})J_{1}(c\widehat{r}_{2})=%
\frac{8\widehat{r}_{1}\widehat{r}_{2}}{(\widehat{r}_{1}^{2}+\widehat{r}%
_{2}^{2})^{3}}\overset{a_{1},a_{2\rightarrow 0}}{\sim }\frac{8\widehat{r}_{1}%
\widehat{r}_{2}}{(\widehat{r}_{1}a_{1}+\widehat{r}_{2}a_{2})^{3}}
\label{K1J1int2}
\end{equation}%
we find that $j(c)=\frac{c^{5}}{8}$\ and the metric function becomes 
\begin{equation}
H=H_{\left( EH\right) ^{2}}(r_{1},r_{2})=1+Q_{M2}\int_{0}^{\infty }dc\{\frac{%
c^{5}}{8}\}\mathcal{R}_{1c}(r_{1})\mathcal{R}_{2c}(r_{2}).
\label{HfinalEH4EH4}
\end{equation}%
As before, changing $c\rightarrow ic$\ in (\ref{eh4xeh4deqR1}) and (\ref%
{eh4xeh4deqR2}) merely interchanges the functions $\mathcal{R}%
_{1}\longleftrightarrow \mathcal{R}_{2}$\ and yields no new solutions.

Next we reduce this metric down to ten dimensions. As with the TN$_{4}$ $%
\otimes $ TN$_{4}$ case above, we can chose to compactify on either $\psi
_{1}$ or $\psi _{2}$, without loss of generality, so again we choose the $%
\psi _{2}$ coordinate. Upon compactification, we get the same NSNS and RR
fields as in the relations (\ref{NSNSEH}) and (\ref{RREH}) in section \ref%
{sec:susyeh4} (with $(w,\theta ,\phi ,g,a)\rightarrow (w_{2},\theta
_{2},\phi _{2},g_{2},a_{2})$), and the metric becomes 
\begin{eqnarray}
ds_{10}^{2} &=&\frac{w_{2}}{2}\{H^{-1/2}g_{2}^{-1/2}\left(
-dt^{2}+dx_{1}^{2}+dx_{2}^{2}\right) +  \notag \\
&+&H^{1/2}g_{2}^{-1/2}\left( ds_{EH(1)}^{2}\right)
+H^{1/2}g_{2}^{1/2}a_{2}^{2}(dw_{2}^{2}+\frac{w_{2}^{2}}{4g_{2}}\left(
d\theta _{2}^{2}+\sin ^{2}(\theta _{2})d\phi _{2}^{2}\right) )\}
\label{ds10EHEH}
\end{eqnarray}%
which describes another D2/D6 system, where $H=H_{\left( EH\right) ^{2}}$. \
In the large $w_{2}$\ limit ($w_{2}=\frac{r_{2}}{a_{2}}$), the metric (\ref%
{ds10EHEH}), reduces to%
\begin{equation}
ds_{10}^{2}=\frac{w_{2}}{2}%
\{-dt^{2}+dx_{1}^{2}+dx_{2}^{2}+ds_{EH(1)}^{2}+dr_{2}^{2}+a_{2}^{2}\frac{%
w_{2}^{2}}{4}\left( d\theta _{2}^{2}+\sin ^{2}(\theta _{2})d\phi
_{2}^{2}\right) \}  \label{ds10EHEHlarger}
\end{equation}%
which is a 10D locally asymptotically flat metric. The Kretchmann invariant
of this spacetime is given by 
\begin{equation}
R_{\mu \nu \rho \sigma }R^{\mu \nu \rho \sigma }=\frac{%
32(48a_{1}^{8}a_{2}^{4}w_{2}^{4}+7r_{1}^{12})}{r_{1}^{12}a_{2}^{4}w_{2}^{6}}%
=(1536\frac{a_{1}^{8}}{r_{1}^{12}})\frac{1}{w_{2}^{2}}+\frac{224}{a_{2}^{4}}%
\frac{1}{w_{2}^{6}}  \label{K1}
\end{equation}%
which vanishes at large $w_{2}$, (though not as rapidly as the case with
only one EH subspace) and all the components of the Riemann tensor in an
orthonormal basis approach zero as $w_{2}\rightarrow \infty $.

This solution also preserves 1/4 of the supersymmetry (and not the 1/8 one
might expect), for the same reason as in section \ref{sec:susytn4xtn4}. The
projection operator (\ref{proj012}) can be used to deal with the gauge
terms, and we're left with two sets of equations of the form (\ref{eh4dpsi}%
), (\ref{eh4dtheta}) and (\ref{eh4dphi}) (with $\psi ,r,\theta ,\phi
\rightarrow \psi _{i},r_{i},\theta _{i},\phi _{i},i=1,2$), which can be
dealt with by two more projection operators; 
\begin{eqnarray}
\Gamma ^{\hat{\psi}_{1}\hat{r}_{1}\hat{\theta}_{1}\hat{\phi}_{1}}\epsilon
&=&-\epsilon  \label{eh4projp1r1th1ph1} \\
\Gamma ^{\hat{\psi}_{2}\hat{r}_{2}\hat{\theta}_{2}\hat{\phi}_{2}}\epsilon
&=&-\epsilon  \label{eh4projp2r2th2ph2}
\end{eqnarray}%
and as before, in eleven dimensions\thinspace\ two of the projection
operators, say (\ref{proj012}) and (\ref{eh4projp1r1th1ph1}), imply the
third.

{\large \ }

\subsection{Taub-NUT$_{4}$ $\otimes $ Eguchi-Hanson}

\label{sec:susytn4xeh4}

Finally, we can also have a Taub-NUT$_{4}$ and an Eguchi-Hanson metric
together embedded into eleven dimensions, giving 
\begin{equation}
ds_{11}^{2}=H(r_{1},r_{2})^{-2/3}\left( -dt^{2}+dx_{1}^{2}+dx_{2}^{2}\right)
+H(r_{1},r_{2})^{1/3}\left( ds_{TN_{4}(1)}^{2}+ds_{EH(2)}^{2}\right) 
\label{metrTNEH}
\end{equation}%
with $ds_{TN_{4}(1)}^{2}$ and $ds_{EH(2)}^{2}$ given by (\ref{dstn4i}) ($i=1$%
) and (\ref{dseh4i}) ($i=2$). This is a solution to the supergravity
equations provided the differential equation 
\begin{eqnarray}
\frac{2}{(r_{1}+2n_{1})}\frac{\partial H}{\partial r_{1}}+\frac{r_{1}}{%
(r_{1}+2n_{1})}\frac{\partial ^{2}H}{\partial r_{1}^{2}}+&&  \notag \\
+\frac{(a_{2}^{4}+3r_{2}^{4})}{r_{2}^{5}}\frac{\partial H}{\partial r_{2}}%
+\frac{(r_{2}-a_{2})(r_{2}+a_{2})(r_{2}^{2}+a_{2}^{2})}{r_{2}^{4}}\frac{%
\partial ^{2}H}{\partial r_{2}^{2}}&=&0  \label{PDETNEH}
\end{eqnarray}%
is satisfied. This is separable using $H(r_{1},r_{2})=1+Q_{M2}R(r_{1})%
\mathcal{R}(r_{2})$. The result is two differential equations, one being (%
\ref{tn4xtn4deqRi}) with ($i=1$) and the other being (\ref{eh4xeh4deqR2}).
The most general solution is given by 
\begin{equation}
H_{A}(r_{1},r_{2})=1+Q_{M2}\int_{0}^{\infty }dck(c)e^{-cr_{1}}\mathcal{U}%
(1+n_{1}c,2,2cr_{1})\mathcal{R}_{2}(r_{2})  \label{HTN4EH4}
\end{equation}%
where we have chosen the separation constant so that the TN solution
exponentially decays. To determine the function $k(c)$ we consider the
special case $n_{1}\rightarrow \infty $\ and $a_{2}\rightarrow 0,$\ where
the radial length in the transverse space to M2 brane is equal to $\sqrt{%
z_{1}^{2}+z_{2}^{2}+\frac{a_{2}^{2}}{4}},$\ yielding%
\begin{equation}
\int_{0}^{\infty }dck(c)\frac{2K_{1}(2c\sqrt{2n_{1}r_{1}})}{c\Gamma (cn_{1})}%
\frac{A_{2c}}{\widehat{r}_{2}}J_{1}(c\widehat{r}_{2})=\frac{\sqrt{2n_{1}r_{1}%
}}{(z_{1}^{2}+z_{2}^{2}+\frac{a_{2}^{2}}{4})^{3}}  \label{HTN4EH4lim}
\end{equation}%
which gives $k(c)=\frac{c^{5}}{16}\Gamma (cn_{1})$, and so in turn%
\begin{equation}
H_{A}(r_{1},r_{2})=1+Q_{M2}\int_{0}^{\infty }dc\frac{c^{5}}{16}\Gamma
(cn_{1})e^{-cr_{1}}\mathcal{U}(1+n_{1}c,2,2cr_{1})\mathcal{R}_{2}(r_{2}).
\label{HTN4EH4A}
\end{equation}%
The alternate case is 
\begin{equation}
H_{B}(r_{1},r_{2})=1+Q_{M2}\int_{0}^{\infty }dcl(c)e^{-icr_{1}}\mathcal{G}%
(1+in_{1}c,2,2icr_{1})\mathcal{R}_{1}(r_{2})  \label{HTN4EH4second}
\end{equation}%
where $\mathcal{R}_{1}(r_{2})$\ is the solution of EH radial differential
equation (\ref{eh4xeh4deqR1}), which decays at large $r_{2},$ (refer to
figure \ref{fig0}). The function $l(c)=\frac{c^{5}}{8}$ by similar methods
and so%
\begin{equation}
H_{B}(r_{1},r_{2})=1+Q_{M2}\int_{0}^{\infty }dc\frac{c^{5}}{8}e^{-icr_{1}}%
\mathcal{G}(1+in_{1}c,2,2icr_{1})\mathcal{R}_{1}(r_{2}).  \label{lc}
\end{equation}

It can of course be shown that this solution preserves 1/4 of the
supersymmetry, by following the same reasoning that shows the Taub-NUT $%
\otimes $ Taub-NUT or Eguchi-Hanson $\otimes $ Eguchi-Hanson solutions
preserve 1/4 of the supersymmetry.

For this solution, we can reduce along either the $\psi_1$ or $\psi_2$
directions, giving two different ten dimensional metrics.

\subsubsection{Case 1: Reduction along $\protect\psi _{1}$}

Reducing along the $\psi _{1}$ direction from the Taub-NUT metric gives the
following fields 
\begin{eqnarray}
\Phi  &=&\frac{3}{4}\ln \left( \frac{H^{1/3}}{\widetilde{f}_{1}}\right)  \\
B_{\mu \nu } &=&0 \\
C_{\phi _{1}} &=&2n_{1}\cos (\theta _{1}) \\
A_{tx_{1}x_{2}} &=&H^{-1}
\end{eqnarray}%
and gives the following metrics 
\begin{eqnarray}
ds_{10}^{2} &=&H^{-1/2}\widetilde{f}_{1}^{-1/2}\left(
-dt^{2}+dx_{1}^{2}+dx_{2}^{2}\right) +H^{1/2}\widetilde{f}%
_{1}^{-1/2}ds_{EH(2)}^{2}+  \notag \\
&+&H^{1/2}\widetilde{f}_{1}^{1/2}\left( dr_{1}^{2}+r_{1}^{2}\left( d\theta
_{1}^{2}+\sin ^{2}(\theta _{1})d\phi _{1}^{2}\right) \right) 
\end{eqnarray}%
which describes two D2/D6 systems (respectively corresponding to $H$\
equalling $H_{A}(r_{1},r_{2})$\ or $H_{B}(r_{1},r_{2})$), where the D2-brane
is localized along the world-volume of the other one.

\subsubsection{Case 2: Reduction along $\protect\psi_2$}

We can also reduce along the $\psi _{2}$ direction from the Eguchi-Hanson
metric, giving fields 
\begin{eqnarray}
\Phi  &=&\frac{3}{4}\ln \left\{ \frac{H^{1/3}w_{2}^{2}}{4g_{2}}\right\}  \\
B_{\mu \nu } &=&0 \\
C_{\phi _{2}} &=&a_{2}\cos (\theta _{2}) \\
A_{tx_{1}x_{2}} &=&\frac{1}{H}
\end{eqnarray}%
and metric 
\begin{eqnarray}
ds_{10}^{2} &=&\frac{w_{2}}{2}\{H^{-1/2}g_{2}^{-1/2}\left(
-dt^{2}+dx_{1}^{2}+dx_{2}^{2}\right) +  \notag \\
&+&H^{1/2}g_{2}^{-1/2}ds_{TN(1)}^{2}+H^{1/2}g_{2}^{1/2}a_{2}^{2}\{dw_{2}^{2}+%
\frac{w_{2}^{2}}{4g_{2}}\left( d\theta _{2}^{2}+\sin ^{2}\theta _{2}d\phi
_{2}^{2}\right) \}\}  \label{ds10TNEH}
\end{eqnarray}%
which describes another pair (for $H$\ equal to either $%
H_{A}(r_{1},r_{2}(w_{2}))$\ or $H_{B}(r_{1},r_{2}(w_{2}))$) of D2/D6 systems
. The world-volume of the D6 brane transverse to D2 is just a TN$_{4}.$ In
the large $w_{2}$\ limit, the metric (\ref{ds10TNEH}), reduces to the metric
(\ref{ds10EHEHlarger}) with $ds_{EH(1)}^{2}\rightarrow ds_{TN_{4}(1)}^{2},$
which is again a 10D locally asymptotically flat metric with the Kretchmann
invariant 
\begin{equation}
R_{\mu \nu \rho \sigma }R^{\mu \nu \rho \sigma }=\frac{A(r_{1},n_{1})}{%
w_{2}^{2}}+\frac{B(r_{1},n_{1},a_{2})}{w_{2}^{6}}  \label{K2}
\end{equation}%
which vanishes at large $w_{2}$ and\ $A(r_{1},n_{1})=\frac{384n_{1}^{2}}{%
(r_{1}+2n_{1})^{6}}$ and 
\begin{equation}
B(r_{1},n_{1},a_{2})=\frac{%
32(7r_{1}^{6}+84n_{1}r_{1}^{5}+420n_{1}^{2}r_{1}^{4}+1120n_{1}^{3}r_{1}^{3}+1680n_{1}^{4}r_{1}^{2}+1344n_{1}^{5}r_{1}+448n_{1}^{6})%
}{(r_{1}+2n_{1})^{6}a_{2}^{4}}.
\end{equation}%
All the components of the Riemann tensor in the orthonormal basis approach
zero at $w_{2}\rightarrow \infty $.

\section{Taub-NUT/Bolt Extended Solutions}

\label{sec:nonsusy}

We consider here solutions that are obtained from generalizations of the
Taub-NUT metric. These are higher-dimensional generalizations of TN$_{4}$
and its bolt-generalizations \cite{TBhigher,awad}. We shall describe these
metrics as we encounter them in turn.

These solutions do not preserve any supersymmetry, but nevertheless exhibit
interesting properties that are qualitatively similar to the previous cases.
Specifically the metric function $H$ behaves the same way near the brane
core and at infinity. For each solution we again find that it is an
integrated product of a decaying function and a damped oscillating function
far from the brane. Near the brane core, the convolution of the two
functions diverges, as for the supersymmetric cases.

We note that a complete list of possible D-brane combinations that give rise
to supersymmetric solutions has been compiled \cite{gau}. These solutions
are obtained by applying the T-duality method to the known supersymmetric
solutions. Although some of the solutions in the following subsections would
seem to be supersymmetric when comparing to this map of solutions, each
metric must be checked explicitly to see if it preserves any supersymmetry.%
\textbf{\ }For example, in subsection \ref{sec:4dtb}, although we find that
we get a $2\perp $ $6(2)$ solution that is on the list, it is not
supersymmetric, as is the situation for all of our solutions in this
section. \ The other system we get is a $2\perp $ $4(2)$ solution which is
not on the list and of course, we also find that this solution is not
supersymmetric.

\subsection{Embedding of Four Dimensional Taub-Bolt Space}

\label{sec:4dtb}

We consider first an M2 brane in the background of Taub-Bolt space given by
the following metric 
\begin{eqnarray}
ds_{11}^{2} &=&H^{-2/3}(y,r)\Big(-dt^{2}+dx_{1}^{2}+dx_{2}^{2}\Big)+  \notag
\\
&+&H^{1/3}(y,r)\Big(dy^{2}+y^{2}[d\alpha ^{2}+\sin ^{2}\alpha (d\beta
^{2}+\sin ^{2}\beta d\gamma ^{2})]+ds_{TB_{4}}^{2}\Big).  \label{ds11tb4}
\end{eqnarray}%
The four-form $F$ has the following non-vanishing components,%
\begin{equation}
F_{tx_{1}x_{2}y}=-\frac{1}{2H}\frac{\partial H}{\partial y}  \label{Ft12ytb4}
\end{equation}%
\begin{equation}
F_{tx_{1}x_{2}r}=-\frac{1}{2H}\frac{\partial H}{\partial r}  \label{Ft12rtb4}
\end{equation}%
and the four-dimensional Taub-Bolt space is given by the metric 
\begin{equation}
ds_{TB_{4}}^{2}=f_{4,Bolt}(r)dr^{2}+f_{4,Bolt}^{-1}(r)\left( d\psi +2n\cos
\theta d\phi \right) ^{2}+(r^{2}-n^{2})\left( d\theta ^{2}+\sin ^{2}\theta
d\phi ^{2}\right)   \label{dstb4}
\end{equation}%
where the function $f_{4}(r)$ is given by 
\begin{equation}
f_{4,Bolt}(r)=\frac{2(r^{2}-n^{2})}{(r-2n)(2r-n)}.  \label{frtb4}
\end{equation}

Note that we have already applied the conditions on $f_{4}(r)$ such that (%
\ref{dstb4}) is a Bolt solution. Briefly, the difference between the NUT and
bolt solutions are as follows. The metric (\ref{dstb4}) is obtained by
fixing the mass parameter in the general metric function of the Taub-NUT
space 
\begin{equation}
f_{4}(r)=\frac{r^{2}-n^{2}}{r^{2}+n^{2}-2mr}  \label{genboltf}
\end{equation}%
so that $f_{4}^{-1}$ vanishes for $r=r_{b}=2n>n$. This occurs if $\ m=\frac{%
5 }{4}n$, \ and the fixed point set of the Killing vector $\partial
/\partial \psi $ is a two-dimensional sphere with radius $\sqrt{3}n$. On the
other hand, if we fix the mass parameter to be $m=n,$ then we have a NUT
solution at $r=r_{n}=n$, where the fixed point set of the Killing vector $%
\partial /\partial \psi $ is zero-dimensional. In this case, the function $%
f_{4}(r)$ reduces to $f_{4,NUT}(r)={\frac{r+n}{r-n}}$ (which is the case we
discussed in section 3). We note that in both cases, $\left. {\frac{1}{%
f_{4,Bolt}(r)}} \right| _{r=r_{b}}=\left. {\frac{1}{f_{4,NUT}(r)}}\right|
_{r=r_{n}}=0$ and $\left. {\frac{d}{dr}}({\frac{1}{f_{4,Bolt}(r)}})\right|
_{r=r_{b}}=\left. {\ \frac{d}{dr}}({\frac{1}{f_{4,NUT}(r)}})\right|
_{r=r_{n}}={\frac{1}{2n}}$.

An equivalent form of the metric (\ref{dstb4}) can be written by shifting $r$
by a constant and scaling $\psi =4n\Psi $, yielding 
\begin{equation}
ds_{TB_{4}}^{2}=\widetilde{f}_{4,Bolt}(r)dr^{2}+16n^{2}\widetilde{f}%
_{4,Bolt}^{-1}(r)\left( d\Psi +\frac{1}{2}\cos \theta d\phi \right)
^{2}+r(r+2n)\left( d\theta ^{2}+\sin ^{2}\theta d\phi ^{2}\right) 
\label{dstb4v2}
\end{equation}%
where 
\begin{equation}
\widetilde{f}_{4,Bolt}(r)=\frac{2r(r+2n)}{(r-n)(2r+n)}  \label{frtb4v2}
\end{equation}%
and the radius $R_{\infty }$ of the circle at $r\rightarrow \infty $, with
line element $R_{\infty }^{2}\left( d\Psi +\frac{1}{2}\cos \theta d\phi
\right) ^{2}$, fibered over base space $S^{2}$ of the Taub-Bolt space, is
given by 
\begin{equation}
R_{\infty }=4n.  \label{Rinfinitytb4}
\end{equation}%
We note that the coordinate $r$ in (\ref{dstb4v2}) ranges over $[n,\infty )$%
, the coordinates $\phi $ and $\Psi $ take values on the interval $[0,2\pi ]$%
, and $\theta $ on the interval $[0,\pi ]$.

The metric (\ref{ds11tb4}) and four-form (\ref{Ft12ytb4},\ref{Ft12rtb4})
satisfy the field equations of $d=11$ supergravity if $H(y,r)$ satisfies the
following equation 
\begin{equation}
(2r^{2}-rn-n^{2})\frac{\partial ^{2}H}{\partial r^{2}}+(4r-n)\frac{\partial H%
}{\partial r}+2r(r+2n)\frac{\partial ^{2}H}{\partial y^{2}}+\frac{6r}{y}%
(r+2n)\frac{\partial H}{\partial y}=0.  \label{laplaceeqtb4}
\end{equation}

\begin{figure}[tbp]
\centering                       
\begin{minipage}[c]{.3\textwidth}
        \centering
        \includegraphics[width=\textwidth]{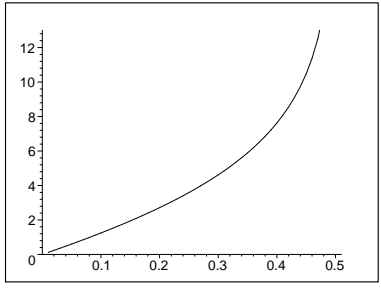}
    \end{minipage}
\caption{ Numerical solution of radial equation (\ref{RequationTB4}) for $%
R/10^{7}$ in Taub-Bolt$_{4}$ case, as a function of $\frac{n}{2r}.$ So for $%
r\approx n$, $R$ diverges and for $r\approx \infty $, it vanishes.}
\label{fig2}
\end{figure}

By separating the coordinates using the relation 
\begin{equation}
H(y,r)=1+Q_{M2}Y(y)R(r)  \label{sepofvartb}
\end{equation}%
we find that the function $Y(y)$ obey equation (\ref{BesselJeqn}), and so 
\begin{equation}
Y(y)\sim \frac{J_{1}\left( cy\right) }{y}.  \label{YTB4}
\end{equation}%
The radial equation is given by 
\begin{equation}
(2r^{2}-n^{2}-rn)R^{\prime \prime }+(4r-n)R^{\prime }-2rc^{2}(r+2n)R=0
\label{RequationTB4}
\end{equation}%
and a typical numerical solution of (\ref{RequationTB4}) versus $\frac{n}{2r}
$ is given in figure \ref{fig2}. The solution $R_{c}(r)$\ logarithmically
diverges at $r=n$\textbf{.}

We note that the most general solution for the metric function is a
superposition of the different functions in the solution (\ref{sepofvartb}),
corresponding to different values of the separation constant $c$. By knowing
the particular numerical solution for $R_{c}(r)$, the most general solution
is 
\begin{equation}
H_{TB_{4}}(y,r)=1+Q_{M2}\int_{0}^{\infty }dcp(c)R_{c}(r)\frac{J_{1}(cy)}{y}.
\label{generalmetricfunctionTB4}
\end{equation}%
By dimensional analysis, $p(c)=p_{0}c^{4}$, where $p_{0}$\ is a constant
that can be absorbed into the definition of $Q_{M2}$. Since the graph of $%
R_{c}(r)$ given in (\ref{fig2}) is qualitatively the same as the
corresponding graph for TN$_{4}$ (both diverge at the brane location, and
vanish at infinity), the behaviour of the metric function (\ref%
{generalmetricfunctionTB4}), is similar to the NUT case (\ref{hashiHyr}).

\begin{figure}[tbp]
\centering                       
\begin{minipage}[c]{.3\textwidth}
        \centering
        \includegraphics[width=\textwidth]{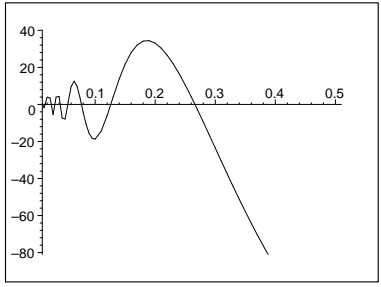}
\end{minipage}
\caption{Numerical solution of damped oscillating radial equation for $%
\widetilde{R}/10$ in Taub-Bolt$_{4}$ case, as a function of $\frac{n}{2r}$.
So for $r\approx n,$ $\widetilde{R}$ diverges and for $r\approx \infty $, it
vanishes.}
\label{figTB4scR}
\end{figure}

The other solution is obtained by changing $c\rightarrow ic$: 
\begin{equation}
\widetilde{H}_{TB_{4}}(y,r)=1+Q_{M2}\int_{0}^{\infty }dc\widetilde{p}(c)%
\widetilde{R_{c}}(r)\frac{K_{1}(cy)}{y}  \label{generalmetricfunctionTB4sc}
\end{equation}%
where $\widetilde{R_{c}}(r)$ is a damped oscillating function. A plot of
this function is given in figure \ref{figTB4scR}. Again by dimensional
analysis, $\widetilde{p}(c)=\widetilde{p}_{0}c^{4}$. The qualitative form of 
$\widetilde{H}_{TB_{4}}$ is similar to that of $\widetilde{H}_{TN}$ in (\ref%
{TN4gfun}), and a plot of $\left( \widetilde{H}_{TB_{4}}(y,r=0)-1\right) $
is the same as the dotted line in figure \ref{figHTN4s}. 

\bigskip

Dimensional reduction of our bolt based solution (\ref{ds11tb4}) in the $%
\Psi $ direction of (\ref{dstb4v2}) gives the type IIA supergravity NSNS
fields: 
\begin{equation}
\begin{array}{rcl}
\Phi  & = & \frac{3}{4}\ln (\frac{\sqrt[3]{H}}{\widetilde{f}_{4,Bolt}(r)})
\\ 
B_{\mu \upsilon } & = & 0%
\end{array}
\label{NSNSTB4}
\end{equation}%
and RR fields $C_{\mu }$ and $A_{\mu \nu \rho }$ have the non-vanishing
components: 
\begin{equation}
\begin{array}{rcl}
C_{\phi } & = & 2n\cos \theta  \\ 
A_{tx_{1}x_{2}} & = & H^{-1}.%
\end{array}
\label{RRfieldsTB4}
\end{equation}%
The \bigskip string-frame 10-dim metric is given by 
\begin{eqnarray}
ds_{10}^{2} &=&\frac{H^{-1/2}}{\sqrt{\widetilde{f}_{4,Bolt}}}\Big(%
-dt^{2}+dx_{1}^{2}+dx_{2}^{2}\Big )+\frac{H^{1/2}}{\sqrt{\widetilde{f}%
_{4,Bolt}}}\Big(dy^{2}+y^{2}d\Omega _{3}^{2}+\widetilde{f}%
_{4,Bolt}dr^{2}+r(r+2n)d\Omega _{2}^{2}\Big)  \notag \\
&=&\frac{H^{-1/2}}{\sqrt{\widetilde{f}_{4,Bolt}}}\Big(%
-dt^{2}+dx_{1}^{2}+dx_{2}^{2}\Big )+\frac{H^{1/2}}{\sqrt{\widetilde{f}%
_{4,Bolt}}}\Big(dy^{2}+y^{2}d\Omega _{3}^{2}\Big)+  \notag \\
&+&H^{1/2}\sqrt{\widetilde{f}_{4,Bolt}}\Big(dr^{\prime 2}+r^{\prime
}(r^{\prime }+\frac{3}{2}n)d\Omega _{2}^{2}\Big)  \label{ds10TB4}
\end{eqnarray}%
where $H$ is equal to either of $H_{TB_{4}}$ or $\widetilde{H}_{TB_{4}}$,
respectively describing distinct D2/D6 brane systems that overlap in two
directions. The function $\widetilde{f}_{4,Bolt}$ is given by relation (\ref%
{frtb4v2}) and we have shifted the coordinate $r$ to $r^{\prime }=r-n$. If
we compare our 10-dim metric with the result obtained in \cite{Hashi}, in
which the Taub-NUT based M-theory was considered, we note that the normal
space to the D2/D6 system is distorted from a perfect sphere (up to the
conformal factor $H^{1/2}\sqrt{\widetilde{f}_{4,Bolt}}$) by the NUT charge.
We have explicitly checked that the above 10-dimensional metric, with the
given dilaton and one form, \textit{is} a solution to the 10-dimensional
supergravity equations of motion.

\bigskip

\subsection{Six dimensional Taub-NUT and Taub-Bolt based M-theory}

\label{sec:6dtntb}

We now demonstrate that M2 branes can be described using higher-dimensional
Taub-NUT/Bolt spaces as backgrounds. Accordingly, we begin by embedding a
six-dimensional Taub-NUT/Bolt space into the eleven-dimensional metric,
given by 
\begin{equation}
ds_{11}^{2}=H^{-2/3}(y,r)(-dt^{2}+dx_{1}^{2}+dx_{2}^{2})+H^{1/3}(y,r)(dy^{2}+y^{2}d\alpha ^{2}+ds_{TN_{6}}^{2})
\label{ds11TN6}
\end{equation}%
and four-form $F$ with non-vanishing components 
\begin{equation}
F_{tx_{1}x_{2}y}=-\frac{1}{2H^{2}}\frac{\partial H}{\partial y}
\label{Ft12yTN6}
\end{equation}%
\begin{equation}
F_{tx_{1}x_{2}r}=-\frac{1}{2H^{2}}\frac{\partial H}{\partial r}
\label{Ft12rTN6}
\end{equation}%
where 
\begin{eqnarray}
ds_{TN_{6}}^{2} &=&f_{6}(r)dr^{2}+f_{6}^{-1}(r)\left( d\psi +2n\cos \theta
_{1}d\phi _{1}+2n\cos \theta _{2}d\phi _{2}\right) ^{2}+  \notag \\
&+&(r^{2}-n^{2})\left( d\theta _{1}^{2}+\sin ^{2}(\theta _{1})d\phi
_{1}^{2}+d\theta _{2}^{2}+\sin ^{2}(\theta _{2})d\phi _{2}^{2}\right) 
\label{dstn6}
\end{eqnarray}%
is the metric of a six dimensional space with NUT charge. The function $f(r)$
is given by 
\begin{equation}
f_{6}(r)=\frac{3(r^{2}-n^{2})^{2}}{r^{4}-6r^{2}n^{2}-6mr-3n^{4}}.  \label{fr}
\end{equation}%
If we fix the mass parameter to be $m=-4n^{3}/3,$ then we have a NUT
solution at $r=r_{n}=n$ and the function $f_{6}(r)$ reduces to 
\begin{equation}
f_{6,NUT}(r)=\frac{3(r+n)^{2}}{(r-n)(r+3n)}.  \label{fr1}
\end{equation}%
On the other hand, to have a bolt solution at $r=r_{b}>n$, we must fix the
mass parameter to be $m=\frac{1}{6}(r_{b}^{3}-6n^{2}r_{b}-\frac{3n^{4}}{r_{b}%
}).$ The bolt is located at $r_{b}=3n$ and the function $\ f_{6}(r)$ reduces
to 
\begin{equation}
f_{6,Bolt}(r)=\frac{3(r-n)^{2}}{(r+n)(r-3n)}.  \label{fr2}
\end{equation}%
In this case, the mass parameter is simply $m=4n^{3}/3$. We note that the
function $f_{6,Bolt}(r)$ can be obtained from $f_{6,NUT}(r)$ by changing $%
n\rightarrow -n$. The locations of the NUT and the bolt are determined by
the relation $\left. {\frac{d}{dr}}({\frac{1}{f_{6,NUT}(r)}})\right|
_{r=r_{n}}=\left. {\frac{d}{dr}}({\frac{1}{f_{6,Bolt}(r)}})\right|
_{r=r_{b}}={\frac{1}{3n}}$ , which is a necessary condition for regularity
of the metric, yielding a consistent period for the coordinate $\psi $.

The radius $R_{\infty }$ of\ circle at $r=\infty ,$ with line element $%
R_{\infty }^{2}\{d\Psi +\frac{1}{3}\cos \theta _{1}d\phi _{1}+\frac{1}{3}%
\cos \theta _{2}d\phi _{2}\}^{2},$ fibered over base space $S^{2}\otimes
S^{2}$ of the Taub-NUT (and Taub-Bolt) space is given by 
\begin{equation}
R_{\infty }=\frac{6}{\sqrt{3}}n=\mp \frac{6}{\sqrt[6]{48}}m^{1/3}.
\label{Rinfinity}
\end{equation}%
In the above equation and in the following, the upper (lower) sign refers to
Taub-NUT (Bolt) space. The coordinates $\phi _{1},\phi _{2}$ and $\Psi $
take values on the interval $\left[ 0,2\pi \right] ,$ while $\theta _{1}$
and $\theta _{2}$ change over $\left[ 0,\pi \right] .$ \ For later
convenience, we can use the following form for the Taub-NUT (Bolt) space 
\begin{eqnarray}
ds_{TN_{6}\pm }^{2} &=&g_{6\pm }(r)dr^{2}+36n^{2}g_{6\pm }^{-1}(r)\left(
d\Psi +\frac{1}{3}\cos \theta _{1}d\phi _{1}+\frac{1}{3}\cos \theta
_{2}d\phi _{2}\right) ^{2}+  \notag \\
&+&r(r\pm 2n)\left( d\theta _{1}^{2}+\sin ^{2}\theta _{1}d\phi
_{1}^{2}+d\theta _{2}^{2}+\sin ^{2}\theta _{2}d\phi ^{2}\right) 
\label{dstn6unus}
\end{eqnarray}%
where 
\begin{equation}
g_{6\pm }(r)=\frac{3(r\pm 2n)^{2}}{r(r\pm 4n)}  \label{gr}
\end{equation}%
and the coordinate $r$ belongs to $[0,\infty )$ for Taub-NUT space and $%
[4n,\infty )$ for Taub-Bolt space (in contrast to (\ref{dstn6}), where the
respective minimum values of the coordinate $r$ are $n$ and $3n$ for the
NUT/Bolt solutions). The metric (\ref{ds11TN6}) and four-form (\ref{Ft12yTN6}%
,\ref{Ft12rTN6}) satisfy the field equations of $d=11$ supergravity,
provided $H(y,r)$ satisfies the following Laplace equation 
\begin{equation}
(r\pm 4n)ry\frac{\partial ^{2}H_{\pm }}{\partial r^{2}}+4y(r\pm 3n)\frac{%
\partial H_{\pm }}{\partial r}+3y(r\pm 2n)^{2}\frac{\partial ^{2}H_{\pm }}{%
\partial y^{2}}+3(r\pm 2n)^{2}\frac{\partial H_{\pm }}{\partial y}=0.
\label{laplaceeq}
\end{equation}%
If we separate the coordinates by 
\begin{equation}
H_{\pm }(y,r)=1+Q_{M2}Y(y)R_{\pm }(r)  \label{sepof var}
\end{equation}%
then the function $Y(y)$, which must remain finite at $y=0$, is given by 
\begin{equation}
Y(y)\approx J_{0}(qy)  \label{Yofy}
\end{equation}%
and after a change of coordinate to $t=r\pm 2n,$ the function $R_{\pm }(t)$
should satisfy 
\begin{equation}
(t^{2}-4n^{2})R_{\pm }^{\prime \prime }(t)+4(t\pm n)R_{\pm }^{\prime
}(t)-3q^{2}t^{2}R_{\pm }(t)=0  \label{Requation}
\end{equation}%
At large distance $r\sim t\rightarrow \infty ,$ the solution of (\ref%
{Requation}) decays according to: 
\begin{equation}
R_{\pm }(t)\approx \frac{(3tq^{2}+\sqrt{3}\left| q\right| )}{t^{3}}e^{-\sqrt{%
3}\left| q\right| t}.  \label{Rinf}
\end{equation}

The metric (\ref{dstn6unus}) for the NUT solution, at small radius $(r\ll n)$
behaves like: 
\begin{eqnarray}
ds_{T\ll 1}^{2} &=&12n^{2}\{dT^{2}+T^{2}(d\Omega _{1}^{2}+\frac{d\Omega
_{2}^{2}+d\Omega _{2}^{^{\prime }2}}{6})\}  \notag \\
&\equiv &d\mathcal{T}^{2}+\mathcal{T}^{2}d\Omega _{5}^{2}  \label{tn6smallr}
\end{eqnarray}%
where $T=\sqrt{(t-2n)/n}$ and $\mathcal{T=}2\sqrt{3}nT$. At small distances $%
r<<n$ (or $\mathcal{T}\rightarrow 0$), equation (\ref{Requation}) has the
solution 
\begin{equation}
R_{+}(\mathcal{T})\approx \frac{1}{\mathcal{T}^{2}}N_{2}(iq\mathcal{T})
\label{Ratsmallr}
\end{equation}%
for the NUT case, where $N_{2}(x)$ is the second order Bessel function of
the second kind. The solution (\ref{Ratsmallr}) diverges as $\frac{1}{%
\mathcal{T}^{4}}$, and decreases with increasing $\mathcal{T}$. \ The other
solution of the equation (\ref{Requation})\textbf{\ }is finite at small
distances.

The divergence of the radial function at small distances is expected
(similar to the TN$_{4}$\ case, which from equation (\ref{RTN4}) diverges
for small $r$, as $\frac{1}{r}$). As in the TN$_{4}$ cases, by a suitable
choice of normalization factor we can find well-defined functions (like
those in figure \ref{figHTN4s}), in terms of the metric functions (\ref%
{generalmetricfunctionTN6}) and (\ref{generalmetricfunctionTN6secondcase}),
that are finite for small values of the radial coordinate $r$.

\begin{figure}[tbp]
\centering                       
\begin{minipage}[c]{.3\textwidth}
        \centering
        \includegraphics[width=\textwidth]{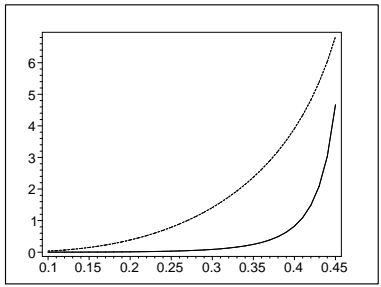}
    \end{minipage}
\caption{Numerical solutions of radial equations (\ref{Requation}) for $%
R_{+}/10^{5}$ and $R_{-}/10^{3}$ in the Taub-NUT$_{6}$ (solid) and Taub-Bolt$%
_{6}$ (dotted) cases, as functions of $\frac{1}{t}.$ So for $r\approx n$ $%
(4n)$ (or $\frac{1}{t}\approx \frac{1}{2n}$ ), $R_{\pm }$ diverge and for $%
r\approx \infty $ (or $\frac{1}{t}\approx 0$), they vanish.}
\label{fig1}
\end{figure}

In the case of the bolt solution, near the brane core $r\sim 4n$ (or $%
t\rightarrow 2n$), the equation (\ref{Requation}), has the solution: 
\begin{equation}
R_{-}(\mathcal{T})\approx N_{0}(iq\mathcal{T})  \label{Rboltatsmallr}
\end{equation}%
where $N_{0}(x)$ is the zeroth order Bessel function of the second kind. The
solution (\ref{Rboltatsmallr}) diverges on the brane as $\ln \mathcal{T}$,
more softly than the NUT case, and decreases for increasing small $\mathcal{T%
}$. A typical numerical solution of (\ref{Requation}) for NUT and bolt cases
versus $\frac{1}{t}$ for $n=1$ is given in figure \ref{fig1}. 

\begin{figure}[tbp]
\centering                       
\begin{minipage}[c]{.3\textwidth}
        \centering
        \includegraphics[width=\textwidth]{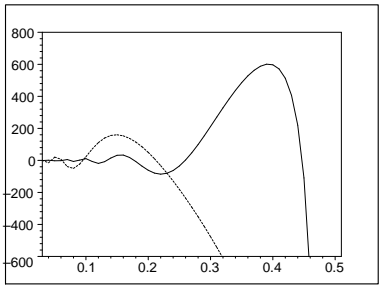}
    \end{minipage}
\caption{Numerical solutions of damped oscillating radial equations for $%
\widetilde{R}_{+}$ and $10^{5}\widetilde{R}_{-}$ in the Taub-NUT$_{6}$
(solid) and Taub-Bolt$_{6}$ (dotted) cases, as functions of $\frac{1}{t}.$
So for $r\approx n$ $(4n)$ (or $\frac{1}{t}\approx \frac{1}{2n}$ ), $%
\widetilde{R}_{\pm }$ diverge and for $r\approx \infty $ (or $\frac{1}{t}%
\approx 0$), they vanish.}
\label{figTN6TB6sc}
\end{figure}

As before, the most general solution is of the form 
\begin{equation}
H_{TN_{6}\pm }(y,r)=1+Q_{M2}\int_{0}^{\infty }dcs_{\pm }(c)R_{c\pm
}(r)J_{0}(cy)  \label{generalmetricfunctionTN6}
\end{equation}%
where by dimensional analysis, the measure function must have the form $%
s_{\pm }(c)=s_{0\pm }c^{5}.$\ The exact value of $s_{0\pm }$\ must be
obtained by looking at near horizon geometry. Two other solutions are
obtained by analytic continuation of $c\rightarrow ic$, yielding 
\begin{equation}
\widetilde{H}_{TN_{6}\pm }(y,r)=1+Q_{M2}\int_{0}^{\infty }dc\widetilde{s}%
_{\pm }(c)\widetilde{R}_{c\pm }(r)K_{0}(cy)
\label{generalmetricfunctionTN6secondcase}
\end{equation}%
where the typical behaviors of the functions $\widetilde{R}_{c\pm }(r)$\ are
plotted in figure \ref{figTN6TB6sc} where we set $n=1$. 

Dimensional reduction of our solution in the $\Psi $ direction of (\ref%
{dstn6unus}) gives the type IIA supergravity NSNS fields: 
\begin{equation}
\begin{array}{rcl}
\Phi  & = & \frac{3}{4}\ln (\frac{\sqrt[3]{H}}{g_{6}(r)}) \\ 
B_{\mu \upsilon } & = & 0%
\end{array}
\label{NSNS}
\end{equation}%
and RR fields $C_{\mu }$ and $A_{\mu \nu \rho }$ have the non-vanishing
components: 
\begin{equation}
\begin{array}{rcl}
C_{\phi _{1}} & = & 2n\cos \theta _{1} \\ 
C_{\phi _{2}} & = & 2n\cos \theta _{2} \\ 
A_{tx_{1}x_{2}} & = & H^{-1}.%
\end{array}
\label{RR6}
\end{equation}

The string-frame 10-dim metric is given by 
\begin{eqnarray}
ds_{10\pm }^{2}
&=&H^{-1/2}g_{6}^{-1/2}(-dt^{2}+dx_{1}^{2}+dx_{2}^{2})+H^{1/2}g_{6}^{-1/2}%
\left( dy^{2}+y^{2}d\alpha ^{2}\right) +  \notag \\
&+&H^{1/2}g_{6}^{1/2}\left( dr^{2}+\frac{r^{2}(r\pm 4n)}{3(r\pm 2n)}\left(
d\Omega _{2}^{2}+d\Omega _{2}^{\prime 2}\right) \right) .  \label{ds10TN6v}
\end{eqnarray}%
The function $H$ is either $H_{TN_{6}\pm }$\ or $\widetilde{H}_{TN_{6}\pm }$
and the function $g_{6}=g_{6\pm }(r)$ given by relation (\ref{gr}). The
10-dim metric describes in each case a D2/D4 brane system, overlapping in
two directions $x_{1}$ and $x_{2}.$

We have explicitly checked that the above 10-dimensional metric, with the
given dilaton and one form, is a solution to the 10-dimensional supergravity
equations of motion. However this solution, in the notation of \cite{gau},
is $2\perp 4(2)$ and does not preserve any supersymmetry. This latter fact
could be obtained by the observation that it cannot be obtained by a
T-dualization of the known supersymmetric D-brane solutions. Note that
T-dualization is a powerful mathematical method for finding new
supersymmetric brane solutions from known supersymmetric solutions. There is
no guarantee that it gives all the supersymmetric brane solutions from a
given theory.

\subsection{Eight dimensional Taub-NUT and Taub-Bolt based M-theory}

\label{sec:8dtntb}

Another case of interest is that of an M2 brane in the background of an
eight-dimensional Taub-NUT/Bolt space. Here we have the following metric%
\begin{equation}
ds_{11}^{2}=H^{-2/3}(r)(-dt^{2}+dx_{1}^{2}+dx_{2}^{2})+H^{1/3}(r)ds_{8}^{2}
\label{ds11tn8}
\end{equation}%
with the 4-form $F$ given by the same non-vanishing components (\ref%
{Ft12yTN6},\ref{Ft12rTN6}). The eight-dimensional space with NUT charge is
given by the metric 
\begin{eqnarray}
ds_{8}^{2} &=&f_{8}(r)dr^{2}+f_{8}^{-1}(r)\left( d\psi +2n\cos \theta
_{1}d\phi _{1}+2n\cos \theta _{2}d\phi _{2}+2n\cos \theta _{3}d\phi
_{3}\right) ^{2}+  \notag \\
&+&(r^{2}-n^{2})\left( d\theta _{1}^{2}+\sin ^{2}\theta _{1}d\phi
_{1}^{2}+d\theta _{2}^{2}+\sin ^{2}\theta _{2}d\phi _{2}^{2}+d\theta
_{3}^{2}+\sin ^{2}\theta _{3}d\phi _{3}^{2}\right)   \label{dstn8}
\end{eqnarray}%
where the function $f_{8}(r)$ is given by 
\begin{equation}
f_{8}(r)=\frac{5(r^{2}-n^{2})^{3}}{r^{6}-5r^{4}n^{2}+15r^{2}n^{4}-10mr+5n^{6}%
}.  \label{frtn8}
\end{equation}

If we fix the mass parameter to be $m=8n^{5}/5,$ then we have a NUT solution
at $r=r_{n}=n$ \ \cite{awad} and the function $f_{8}(r)$ reduces to 
\begin{equation}
f_{8N}(r)=\frac{5(r+n)^{3}}{r^{3}+3nr^{2}+n^{2}r-5n^{3}}.  \label{frn81}
\end{equation}%
On the other hand, to have a bolt solution at $r=r_{b}>n$, we must fix the
mass parameter to be $m=\frac{1}{10}(r_{b}^{5}-5n^{2}r_{b}^{3}+15n^{4}r_{b}+%
\frac{5n^{6}}{r_{b}}).$ The bolt is located at $r_{b}=4n$ and the function $%
f_{8}(r)$ reduces to 
\begin{equation}
f_{8B}(r)=\frac{20(r^{2}-n^{2})^{3}}{%
4r^{6}-20n^{2}r^{4}+60n^{4}r^{2}-3061n^{5}r+20n^{6}}  \label{frn82}
\end{equation}%
where the mass parameter simplifies to $m=3061n^{5}/40$.

The radius $R_{\infty }$ of\ circle at $r=\infty $, with line element $%
R_{\infty }^{2}\{d\Psi +\frac{1}{4}\cos \theta _{1}d\phi _{1}+\frac{1}{4}%
\cos \theta _{2}d\phi _{2}+\frac{1}{4}\cos \theta _{3}d\phi _{3}\}^{2}$,
fibered over base space $S^{2}\otimes S^{2}\otimes S^{2}$ of the space (\ref%
{dstn8}) is given by 
\begin{equation}
R_{\infty }=\frac{8}{\sqrt{5}}n  \label{Rinfinitytn8}
\end{equation}%
which for Taub-NUT space is equal to $\frac{\sqrt[5]{4096}}{\sqrt[10]{125}}%
m^{1/5}$ and for Taub-Bolt space is equal to $\frac{8\sqrt[5]{40}}{\sqrt{5}%
\sqrt[5]{3061}}m^{1/5}.$ The coordinates $\phi _{i}$ and $\Psi $ take values
on the interval $[0,2\pi ]$, while the $\theta _{i}$ range over $[0,\pi ]$.
For later convenience, we can use the following form for the Taub-NUT (Bolt)
space 
\begin{eqnarray}
ds_{8}^{2} &=&g_{8}(r)dr^{2}+64n^{2}g_{8}^{-1}(r)\{d\Psi +\frac{1}{4}\cos
\theta _{1}d\phi _{1}+\frac{1}{4}\cos \theta _{2}d\phi _{2}+\frac{1}{4}\cos
\theta _{3}d\phi _{3}\}^{2}+  \notag \\
&+&r(r+2n)\{d\theta _{1}^{2}+\sin ^{2}\theta _{1}d\phi _{1}^{2}+d\theta
_{2}^{2}+\sin ^{2}\theta _{2}d\phi _{2}^{2}+d\theta _{3}^{2}+\sin ^{2}\theta
_{3}d\phi _{3}^{2}\}  \label{dstn8v2}
\end{eqnarray}%
where the function $g_{8}(r)$ for Taub-NUT space is 
\begin{equation}
g_{8N}(r)=\frac{5(r+2n)^{3}}{r(r^{2}+6nr+10n^{2})}  \label{grtn8}
\end{equation}%
and for Taub-Bolt space is given by 
\begin{equation}
g_{8B}(r)=\frac{5r^{3}(r+2n)^{3}}{r^{6}+6nr^{5}+10n^{2}r^{4}-\frac{2997}{4}%
n^{5}(r+n)}  \label{grtb8}
\end{equation}%
and the coordinate $r$ belongs to $[0,\infty )$ for the NUT solution and to $%
[3n,\infty )$ for the bolt solution. Note that in (\ref{dstn8}), the minimum
value of the coordinate $r$ is $n$ for the NUT solution and $4n$ for the
bolt solution.

\subsection{Taub-NUT case}

In this subsection, we consider the NUT solution. The metric (\ref{ds11tn8})
with the NUT solution metric function (\ref{grtn8}) and four-form (\ref%
{Ft12yTN6},\ref{Ft12rTN6}) satisfy the field equations of $d=11$
supergravity, if $H(r)$ satisfies to the following Laplace equation

\begin{equation}
r(r^{2}+6rn+10n^{2})\frac{\partial ^{2}H}{\partial r^{2}}%
+2(3r^{2}+15rn+20n^{2})\frac{\partial H}{\partial r}=0
\label{laplaceeqtn8fonlyofr}
\end{equation}%
which has an exact solution 
\begin{equation}
H(r)=\frac{1}{30n^{2}r^{3}}-\frac{3}{100n^{3}r^{2}}+\frac{13}{500n^{4}r}+%
\frac{12\ln \frac{r^{2}}{r^{2}+6rn+10n^{2}}-7\{\tan ^{-1}(\frac{r}{n}+3)-%
\frac{\pi }{2}\}}{2500n^{5}}  \label{Hofrexact}
\end{equation}%
that behaves appropriately at small and large distances. For large values of 
$r,$ it vanishes as%
\begin{equation}
H(r)\sim \frac{1}{r^{5}}  \label{HTN8larger}
\end{equation}%
\bigskip and for small values of $r,$ it diverges as 
\begin{equation}
H(r)\sim \frac{1}{r^{3}}.  \label{HTN8smallr}
\end{equation}

In this case, the dimensional reduction of our solution in the $\Psi $
direction of (\ref{dstn8v2}) gives the type IIA supergravity NSNS fields: 
\begin{equation}
\begin{array}{rcl}
\Phi  & = & \frac{3}{4}\ln (\frac{\sqrt[3]{H}}{g_{8}(r)}) \\ 
B_{\mu \upsilon } & = & 0%
\end{array}
\label{NSNSTN8}
\end{equation}%
and the RR fields $C_{\mu }$ and $A_{\mu \nu \rho }$ have the non-vanishing
components: 
\begin{equation}
\begin{array}{c}
C_{\phi _{1}}=2n\cos \theta _{1} \\ 
C_{\phi _{2}}=2n\cos \theta _{2} \\ 
C_{\phi _{3}}=2n\cos \theta _{3} \\ 
A_{tx_{1}x_{2}}=H^{-1}.%
\end{array}
\label{RRfieldsTN8}
\end{equation}%
The string-frame 10-dim metric 
\begin{equation}
ds_{10}^{2}=\frac{H^{-1/2}}{\sqrt{g_{8}(r)}}(-dt^{2}+dx_{1}^{2}+dx_{2}^{2})+%
\frac{H^{1/2}}{\sqrt{g_{8}(r)}}\{g_{8}(r)dr^{2}+r(r+2n)(d\Omega
_{2}^{2}+d\Omega _{2}^{\prime 2}+d\Omega _{2}^{\prime \prime 2})\}
\label{ds10TN8}
\end{equation}%
describes a D2 brane localized at a point of the 7-dimensional background
space where the function $g_{8}(r)$ is given by relation (\ref{grtn8}). We
have explicitly checked that the above 10-dimensional metric, with the given
dilaton and one form, is a solution to the 10-dimensional supergravity
equations of motion.

\subsection{Taub-Bolt case}

\begin{figure}[tbp]
\centering                       
\begin{minipage}[c]{.3\textwidth}
        \centering
        \includegraphics[width=\textwidth]{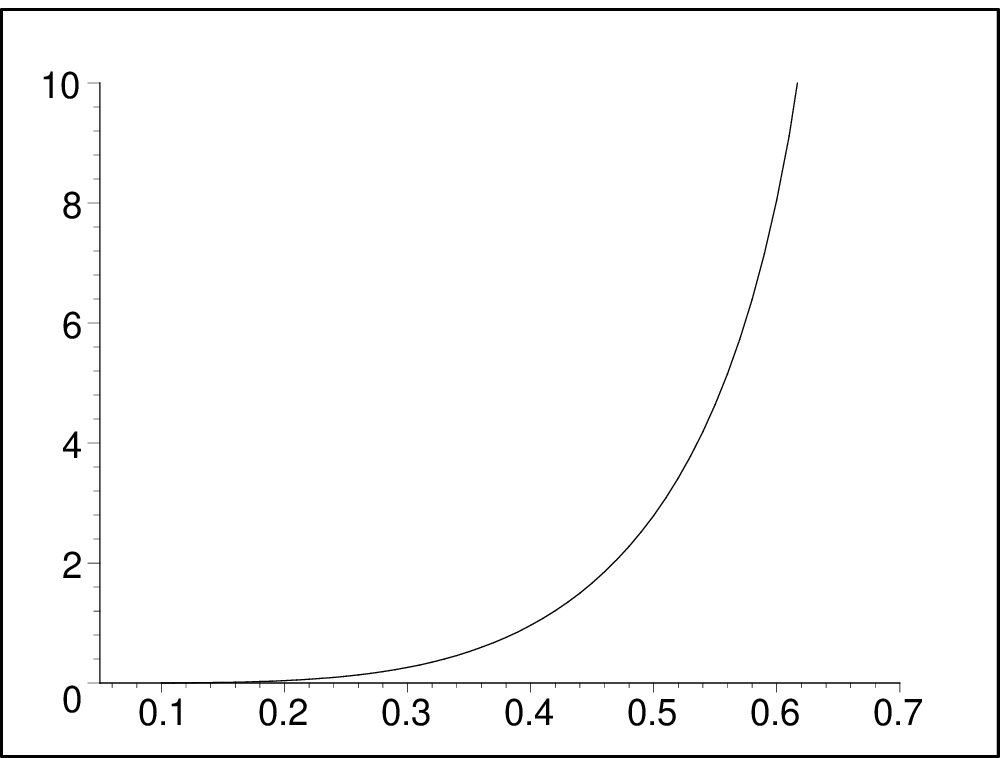}
    \end{minipage}
\caption{ Numerical solution of Laplace equation (\ref{laplaceeqtb8}) for $%
H/10^{25}$ in Taub-Bolt$_{8}$, as a function of $\frac{2n}{r}.$ So for $%
r\approx 3n$, $H$ diverges and for $r\approx \infty ,$ vanishes.}
\label{fig5}
\end{figure}

Now, we consider the Taub-Bolt case. The metric (\ref{ds11tn8}) with the
bolt solution metric function (\ref{grtb8}) and four-form (\ref{Ft12yTN6}, %
\ref{Ft12rTN6}) satisfy the field equations of $d=11$ supergravity, if $H(r)$
satisfies to the following Laplace equation 
\begin{equation}
\{4r^{6}+24nr^{5}+40n^{2}r^{4}-2997n^{5}(r+n)\}H^{\prime \prime
}+(24r^{5}+120nr^{4}+160n^{2}r^{3}-2997n^{5})H^{\prime }=0
\label{laplaceeqtb8}
\end{equation}%
which has the solution%
\begin{equation}
H=\int \frac{d\widetilde{r}}{(16875n^{5}+13500\widetilde{r}n^{4}+4800%
\widetilde{r}^{2}n^{3}+940\widetilde{r}^{3}n^{2}+96\widetilde{r}^{4}n+4%
\widetilde{r}^{5})\widetilde{r}}  \label{HTB8}
\end{equation}%
where we have set \ $r\rightarrow \widetilde{r}=r-3n$\ so that $\widetilde{r}%
=0$\ is the location of the bolt. The function $H$\ diverges logarithmically
near the bolt and vanishes at infinity as $\frac{1}{\widetilde{r}^{5}}$%
\textbf{. }A numerical solution of the equation (\ref{laplaceeqtb8}) is
shown in figure \ref{fig5}. 
The type IIA supergravity NSNS, RR fields and 10-dimensional string metric
have the same form as (\ref{NSNSTN8}), (\ref{RRfieldsTN8}) and (\ref{ds10TN8}
) where the function $g_{8}(r)$ is given by relation (\ref{grtb8}).

\section{Decoupling limits}

In this section we wish to discuss the decoupling limits for the various
solutions we have presented above. The specifics of calculating the
decoupling limit are shown in detail elsewhere (see for example \cite%
{DecouplingLim}), so we will only provide a brief outline here. The process
is the same for all cases, so we will also only provide specific examples of
a few of the solutions above.

At low energies, the dynamics of the D2 brane decouple from the bulk, with
the region close to the D6 brane corresponding to a range of energy scales
governed by the IR fixed point \cite{DecouplingLim}. For D2 branes localized
on D6 branes, this corresponds in the field theory to a vanishing mass for
the fundamental hyper-multiplets. Near the D2 brane horizon ($H\gg 1$), the
field theory limit is given by 
\begin{equation}
g_{YM2}^{2}=g_{s}\ell _{s}^{-1}=\text{fixed.}  \label{gymFTlimit}
\end{equation}%
In this limit the gauge couplings in the bulk go to zero, so the dynamics
there decouple. In each of our cases above, the radial coordinates are also
scaled such that 
\begin{equation}
Y=\frac{y}{\ell _{s}^{2}}~~,~~~~U_{i}=\frac{r_{i}}{\ell _{s}^{2}}
\label{YUdecoupling1}
\end{equation}%
are fixed (where $Y$ and $U_{i}$, $i=1,2$ are used where appropriate). As an
example we note that this will change the harmonic function of the D6 brane
in the TN$_{4}$ $\otimes $ TN$_{4}$ case to the following (recall that the
asymptotic radius of the 11th dimension is $R_{\infty }=4n_{2}=g_{s}\ell _{s}
$) 
\begin{equation}
f_{2}(r_{2})=\left( 1+\frac{2n_{2}}{r_{2}}\right) =\left( 1+\frac{g_{s}\ell
_{s}}{2r_{2}}\right) =\left( 1+\frac{g_{s}}{2\ell _{s}U_{2}}\right) =\left(
1+\frac{g_{YM2}^{2}N_{6}}{2U_{2}}\right) =f(U_{2})  \label{F2}
\end{equation}%
(generalizing to the case of $N_{6}$ D6 branes as was done in \cite{Hashi},
giving the factor of $N_{6}$ in the final line above - this function is of
course the same as the harmonic function of the D6 brane for the TN$_{4}$
case from \cite{Hashi}, as is to be expected). For the Taub-NUT case, $f(U)$
is given by equation (\ref{F2}), whereas for the EH metric it is 
\begin{equation}
g(r)\rightarrow \left( 1-\frac{A^{4}}{U^{4}}\right) =g(U)  \label{EHfU}
\end{equation}%
where $a$ has been rescaled to $a=A\ell _{s}^{-2}$.

All of the D2 harmonic functions from the above solutions (both
supersymmetric and non-supersymmetric) can be shown to scale as $H(Y,U)=\ell
_{s}^{-4}h(Y,U)$. This form causes the D2-brane to warp the ALE region and
the asymptotically flat region of the D6-brane geometry. The $h(Y,U)$'s are
easily calculated; as an example, the TN$_{4}$ $\otimes $ TN$_{4}$ function
is given by 
\begin{eqnarray}
h_{\left( TN\right) ^{2}}(U_{1},U_{2}) &=&\frac{1}{2}\pi
^{2}N_{2}g_{YM2}^{2}\int_{0}^{\infty }dPP^{5}\Gamma \left( \frac{Pg_{YM2}^{2}%
}{4}\right) e^{-iPU_{1}}\times   \notag \\
&&\times \mathcal{G}_{1}\left( 1+iPm_{1},2,2iPU_{1}\right) e^{-PU_{2}}%
\mathcal{U}\left( 1+\frac{Pg_{YM2}^{2}}{4},2,2PU_{2}\right)   \label{hTN2}
\end{eqnarray}%
where the $U_{i}$ in (\ref{YUdecoupling1}) with $i=1$ and $2$ have been
used, as well as rescaling the $n_{1}=m_{1}\ell _{s}^{2}$, $c=P/\ell _{s}^{2}
$. Even when full analytic forms of $H(y,r)$ are not available, we can show
that $H(Y,U)=\ell _{s}^{-4}h(Y,U)$ in the decoupling limit, due to the
general forms of $H(y,r)$ we obtained above.

For the $H(Y,U)$ functions, we use $\ell _{p}=g_{s}^{1/3}\ell _{s}$ to
rewrite 
\begin{equation}
Q_{M2}=32\pi ^{2}N_{2}\ell _{p}^{6}=32\pi ^{2}N_{2}g_{YM2}^{2}\ell _{s}^{8}.
\label{Qm2value}
\end{equation}%
The respective supersymmetric metrics are then given by appropriate
insertion of either (\ref{F2}) or (\ref{EHfU}) along with the relevant $%
h(Y,U)$ into the metrics (\ref{ds10TN4}), (\ref{dsEH10}), (\ref{ds10tnXtn}),
(\ref{ds10EHEH}). For example, for TN$_{4}$, the metric (\ref{ds10TN4})
scales as \cite{Hashi}: 
\begin{eqnarray}
\frac{ds_{10}^{2}}{\ell _{s}^{2}} &=&h(Y,U)^{-1/2}f(U)^{-1/2}\left(
-dt^{2}+dx_{1}^{2}+dx_{2}^{2}\right) +  \notag \\
&+&h(Y,U)^{1/2}f(U)^{-1/2}\left( dY^{2}+Y^{2}d\Omega _{3}^{2}\right) + 
\notag \\
&+&h(Y,U)^{1/2}f(U)^{1/2}\left( dU^{2}+U^{2}d\Omega _{2}^{2}\right) 
\end{eqnarray}%
and there is only an overall normalization factor of $\ell _{s}^{2}$ in the
above metric. This is the expected result for a solution that is a
supergravity dual of a QFT. As all of the other supersymmetric cases are
qualitatively the same, we won't write them out here.

For the non-supersymmetric cases, we take the same limits (\ref{gymFTlimit}%
), (\ref{YUdecoupling1}), but of course the metric functions will differ. As
an example, for the 4-dimensional Taub-Bolt case, this will change the
harmonic function of the D6 brane to the following 
\begin{equation}
\widetilde{f_{4}}(U)=\frac{2r(r+2n)}{(r-n)(2r+n)}=\frac{16U(g_{YM2}^{2}+2U)}{%
(8U+g_{YM2}^{2})(4U-g_{YM2}^{2})}
\end{equation}%
where $R_{\infty }=4n=g_{s}\ell _{s}$ has been used. The metric functions of
the non-D2 branes in the TN$_{6},$TB$_{6}$ and TN$_{8},$TB$_{8}$ cases will
be altered by the same steps. Since an analytic function for $R(r)$ cannot
be found in any of the non-susy cases, the transformations of the D2
functions $H(y,r)$ cannot be found explicitly. However, using the general
forms (\ref{generalmetricfunctionTB4}), (\ref{generalmetricfunctionTN6}) we
can still show, for example, that 
\begin{equation}
H_{TB4}(y,r)\rightarrow Q_{M2}\int \frac{dP}{\ell _{s}^{2}}p_{0}\frac{P^{4}}{%
\ell _{s}^{8}}\frac{R_{c}(U)J_{1}(PY)}{Y\ell _{s}^{2}}=\frac{1}{\ell _{s}^{4}%
}h_{TB_{4}}(Y,U)  \label{hTB4}
\end{equation}%
where (\ref{Qm2value}), (\ref{YUdecoupling1}) and the rescaling $c=P\ell
_{s}^{-2}$ have been used. The TB$_{4}$ metric will then become 
\begin{eqnarray}
\frac{ds^{2}}{\ell _{s}^{2}} &=&\frac{1}{\sqrt{\widetilde{f_{4}}%
(U)h_{TB_{4}}(Y,U)}}(-dt^{2}+dx_{1}^{2}+dx_{2}^{2})+\frac{\sqrt{%
h_{TB_{4}}(Y,U)}}{\sqrt{\widetilde{f_{4}}(U)}}(dY^{2}+Y^{2}d\Omega _{3}^{2})+
\notag \\
&+&\sqrt{\widetilde{f_{4}}(U)h_{TB_{4}}(Y,U)}\left( dU^{2}+U\left( U+\frac{%
3g_{YM}^{2}}{8}\right) d\Omega _{2}^{2}\right)   \label{tb4decoupled}
\end{eqnarray}%
and the TN$_{6}$/TB$_{6}$ metric in the decoupling limit will be given by 
\begin{eqnarray}
\frac{ds_{10}^{2}}{\ell _{s}^{2}} &=&\frac{1}{\sqrt{h_{TN_{6}}(Y,U)g_{6}(U)}}%
\left( -dt^{2}+dx_{1}^{2}+dx_{2}^{2}\right) +\sqrt{\frac{h_{TN_{6}}(Y,U)}{%
g_{6}(U)}}\left( dY^{2}+Y^{2}d\alpha ^{2}\right) +  \notag \\
&+&\sqrt{h_{TN_{6}}(Y,U)g_{6}(U)}\left( dU^{2}+\frac{U^{2}(\sqrt{3}U\pm
2g_{YM}^{2})}{3(\sqrt{3}U\pm g_{YM}^{2})}\left( d\Omega ^{2}+d\Omega
^{\prime 2}\right) \right) 
\end{eqnarray}%
where again we note that the only dependence on the string scale $\ell _{s}$
is in an overall normalization factor. There is of course no decoupling
occurring in the eight dimensional Taub-NUT/Bolt cases.

\section{Conclusions}

By embedding different possible spaces with NUT charge and/or self-dual
curvature, we have found a rich and interesting new class of brane solutions
to D=11 supergravity. These exact solutions are new M2 brane metrics, and
are presented in equations (\ref{HM2secondcase}), (\ref{HfinalEH4}), (\ref%
{HEH4sc}), (\ref{HTN4TN4}), (\ref{HfinalEH4EH4}), (\ref{HTN4EH4A}), (\ref{lc}%
), (\ref{generalmetricfunctionTB4}), (\ref{generalmetricfunctionTB4sc}), (%
\ref{generalmetricfunctionTN6}), (\ref{generalmetricfunctionTN6secondcase}),
(\ref{Hofrexact}) and (\ref{HTB8}) which are the main results of this paper.
The common feature of all of these solutions is that the brane function is a
convolution of an exponentially decaying `radial' function with a damped
oscillating one. The `radial' functions vanish far from the M2 brane and
diverge near the brane core.

Dimensional reduction to 10 dimensions gives us different D2/D6 systems
(with metric functions) which in all the cases with a combination of 4
dimensional Taub-NUT and Eguchi-Hanson spaces (which have self-dual Riemann
curvature), the configurations preserve 1/4 of the supersymmetry and yield
metrics with acceptable asymptotic behaviour. In all other cases --
involving 4 dimensional Taub-Bolt and higher dimensional Taub-NUT/Bolt --
the D-brane system is not supersymmetric. However the general functional
structure of the branes is qualitatively the same as for the supersymmetric
cases: the $r$-dependent parts of the metric functions diverge for small $r$%
\ and fall off rapidly for large $r$, whereas the $y$-dependent parts of the
metric functions approach a finite value for small $y$\ and vanish at large $%
y.$

Finally we considered the decoupling limit of our solutions. In the case of
embedded TN$_{4}$, when the D2 brane decouples from the bulk, the theory on
the brane is 3 dimensional $\mathcal{N}=4$ $SU($N$_{2})$ super Yang-Mills
(with eight supersymmetries) coupled to N$_{6}$\ massless hypermultiplets %
\cite{pelc}. This point is obtained from dual field theory and since some of
our solutions preserve the same amount of supersymmetry, a similar dual
field description should be attainable. \bigskip

{\Large Acknowledgments}

We would like to thank J. Gauntlett for helpful discussions. This work was
supported by the Natural Sciences and Engineering Research Council of Canada.

\appendix

\section{Supersymmetries of the TN and EH solutions}

\label{sec:susycalc}

\bigskip We demonstrate in this appendix the supersymmetry of our solutions
that embed one (or both) of TN$_{4}$ and EH.

We begin by noting another useful representation of the algebra (\ref%
{CliffordAlg}), given by the following 32 dimensional representation \cite%
{gauntlettetal3} 
\begin{equation}
\begin{array}{c}
\Gamma _{i}=\left[ 
\begin{array}{cc}
0 & -\widetilde{\Gamma }_{i} \\ 
\widetilde{\Gamma }_{i} & 0%
\end{array}%
\right] ~~~(i=1\ldots 8) \\ 
\Gamma _{9}=\left[ 
\begin{array}{cc}
1_{16} & 0 \\ 
0 & -1_{16}%
\end{array}%
\right] \\ 
\Gamma _{\#}=\left[ 
\begin{array}{cc}
0 & 1_{16} \\ 
1_{16} & 0%
\end{array}%
\right] \\ 
\Gamma _{0}=-\Gamma _{123456789\#}%
\end{array}
\label{Gammas}
\end{equation}%
{\large \ }where $\widetilde{\Gamma }_{i}$, the 16 dimensional
representation of the Clifford algebra in eight dimension, are given by%
\begin{eqnarray}
\widetilde{\Gamma }_{i} &=&\left[ 
\begin{array}{cc}
0 & L_{i} \\ 
L_{i} & 0%
\end{array}%
\right] ~~~(i=1\ldots 7)  \label{Gammatildas} \\
\widetilde{\Gamma _{8}} &=&\left[ 
\begin{array}{cc}
0 & -1_{8} \\ 
1_{8} & 0%
\end{array}%
\right]
\end{eqnarray}%
{\large \ }in terms of $L_{i},$\ the left multiplication by the imaginary
octonions on the octonions. The imaginary unit octonions satisfy the
following relationship 
\begin{equation}
e_{i}\cdot e_{j}=-\delta _{ij}+c_{ijk}e_{k}  \label{unitoct}
\end{equation}%
where $c_{ijk}$\ is totally skew symmetric and its non-vanishing components
are given by

\begin{equation}
c_{124}=c_{137}=c_{156}=c_{235}=c_{267}=c_{346}=c_{457}=1.
\label{nonvanishingcompo}
\end{equation}%
We take the $L_{i}$ to be the matrices such that the relation (\ref{unitoct}%
) holds. In other words, given a vector $v=\left( v_{0},v_{i}\right) $ in $%
\mathbb{R}^{8}$, we write $\hat{v}=v_{0}+v_{j}e_{j}$, where the effect of
left multiplication is $e_{i}\left( \hat{v}\right)
=v_{0}e_{i}-v_{i}+c_{ijk}v_{j}e_{k}$ , we then construct the $8\times 8$
matrix $\left( L_{i}\right) _{AB}$ by requiring $e_{i}\left( \hat{v}\right)
=\left( L_{i}\right) _{AB}e_{A}v_{B}$, where $A,B=0,1,\ldots 7$. With this
in mind, we obtain considerable simplification of the Killing spinor
equation (\ref{genkspinor}).

Two relations that are useful in solving for the Killing spinors are 
\begin{eqnarray}
e^{A}Be^{-A} &=&B+\left[ A,B\right] +\frac{1}{2!}\left[ A,\left[ A,B\right] %
\right] +\ldots   \label{BCHid} \\
\left[ \Gamma ^{ab},\Gamma ^{cd}\right]  &=&2\eta ^{ad}\Gamma ^{bc}+2\eta
^{bc}\Gamma ^{ad}-2\eta ^{ac}\Gamma ^{bd}-2\eta ^{bd}\Gamma ^{ac}.
\label{GammId1}
\end{eqnarray}%
Eq. (\ref{BCHid}) is of course the Baker-Campbell-Hausdorff identity, and (%
\ref{GammId1}) is derivable from (\ref{CliffordAlg}) and (\ref{Gamma1top}),
for $p=2$.

Consider first all of our supersymmetry preserving solutions. Using (\ref%
{genkspinor}), one finds 
\begin{equation}
\left( 1+\Gamma ^{\hat{t}\hat{x}_{1}\hat{x}_{2}}\right) \epsilon =0
\label{proj012}
\end{equation}%
(where hats over coordinates denote tangent space indices), and so at most
half the supersymmetry is preserved due to the presence of the brane.

Now, consider an embedding of TN$_{4}$. The remaining equations from (\ref%
{genkspinor}), arising from the left-over terms from $\partial _{M}\epsilon +%
\textstyle\frac{1}{4}\omega _{Mab}\Gamma ^{ab}\epsilon $\ portion, are 
\begin{eqnarray}
\partial _{\alpha _{1}}\epsilon -\frac{1}{2}\Gamma ^{\hat{y}\hat{\alpha}%
_{1}}\epsilon  &=&0  \label{tn4da1} \\
\partial _{\alpha _{2}}\epsilon -\frac{\sin (\alpha _{1})}{2}\Gamma ^{\hat{y}%
\hat{\alpha}_{2}}\epsilon -\frac{\cos (\alpha _{1})}{2}\Gamma ^{\hat{\alpha}%
_{1}\hat{\alpha}_{2}}\epsilon  &=&0  \label{tn4da2} \\
\partial _{\alpha _{3}}\epsilon -\frac{\sin (\alpha _{1})\sin (\alpha _{2})}{%
2}\Gamma ^{\hat{y}\hat{\alpha}_{3}}\epsilon -\frac{\sin (\alpha _{2})\cos
(\alpha _{1})}{2}\Gamma ^{\hat{\alpha}_{1}\hat{\alpha}_{3}}\epsilon -\frac{%
\cos (\alpha _{2})}{2}\Gamma ^{\hat{\alpha}_{2}\hat{\alpha}_{3}}\epsilon 
&=&0  \label{tn4da3} \\
\partial _{\psi }\epsilon +\frac{n}{2(r+n)^{2}}\left[ \Gamma ^{\hat{\psi}%
\hat{r}}+\Gamma ^{\hat{\theta}\hat{\phi}}\right] \epsilon  &=&0
\label{tn4dpsi} \\
\partial _{\theta }\epsilon +\frac{n}{2(r+n)}\Gamma ^{\hat{\psi}\hat{\phi}%
}\epsilon -\frac{r}{2(r+n)}\Gamma ^{\hat{r}\hat{\theta}}\epsilon  &=&0
\label{tn4dtheta} \\
\partial _{\phi }\epsilon +\frac{n^{2}\cos (\theta )}{(r+n)^{2}}\left[
\Gamma ^{\hat{\psi}\hat{r}}+\Gamma ^{\hat{\theta}\hat{\phi}}\right] \epsilon
-\frac{n\sin (\theta )}{2(r+n)}\Gamma ^{\hat{\psi}\hat{\theta}}\epsilon -%
\frac{r\sin (\theta )}{2(r+n)}\Gamma ^{\hat{r}\hat{\phi}}\epsilon -\frac{%
\cos (\theta )}{2}\Gamma ^{\hat{\theta}\hat{\phi}}\epsilon  &=&0.~~  \notag
\\
&&  \label{tn4dphi}
\end{eqnarray}%
Equations (\ref{tn4da1}), (\ref{tn4da2}) and (\ref{tn4da3}) are solvable by
use of (\ref{GammId1}) and (\ref{BCHid}) to give the following Lorentz
rotation 
\begin{equation}
\epsilon =\exp \left\{ \frac{\alpha _{1}}{2}\Gamma ^{\hat{y}\hat{\alpha}%
_{1}}\right\} \exp \left\{ \frac{\alpha _{2}}{2}\Gamma ^{\hat{\alpha}_{1}%
\hat{\alpha}_{2}}\right\} \exp \left\{ \frac{\alpha _{3}}{2}\Gamma ^{\hat{%
\alpha}_{2}\hat{\alpha}_{3}}\right\} \tilde{\epsilon}.  \label{tn4LorRota123}
\end{equation}%
Now note that (\ref{tn4dpsi}) can be solved for by using the projection
operator 
\begin{equation}
\Gamma ^{\hat{\psi}\hat{r}\hat{\theta}\hat{\phi}}\epsilon =\epsilon 
\label{projprthph}
\end{equation}%
eliminating another half of the supersymmetry provided $\epsilon $\ is
independent of $\psi $. With this projection operator, (\ref{tn4dtheta}) and
(\ref{tn4dphi}) can be solved, giving 
\begin{equation}
\epsilon =\exp \left\{ -\frac{\theta }{2}\Gamma ^{\hat{\psi}\hat{\phi}%
}\right\} \exp \left\{ \frac{\phi }{2}\Gamma ^{\hat{\theta}\hat{\phi}%
}\right\} \tilde{\epsilon}.  \label{tn4epsilon}
\end{equation}

Turning next to the EH metric, the remaining terms from the Killing spinor
equation (\ref{genkspinor}) give (\ref{tn4da1}), (\ref{tn4da2}) and (\ref%
{tn4da3}), which have the same solution as above, and three more equations 
\begin{eqnarray}
\partial _{\psi }\epsilon +\frac{(r^{4}+a^{4})}{4r^{4}}\Gamma ^{\hat{\psi}%
\hat{r}}\epsilon +\frac{(r^{4}-a^{4})}{4r^{4}}\Gamma ^{\hat{\theta}\hat{\phi}%
}\epsilon  &=&0  \label{eh4dpsi} \\
\partial _{\theta }\epsilon -\frac{1}{4g^{1/2}}\Gamma ^{\hat{r}\hat{\theta}%
}\epsilon +\frac{1}{4g^{1/2}}\Gamma ^{\hat{\psi}\hat{\phi}}\epsilon  &=&0
\label{eh4dtheta} \\
\partial _{\phi }\epsilon -\frac{\sin (\theta )}{4g^{1/2}}\Gamma ^{\hat{\psi}%
\hat{\theta}}\epsilon -\frac{\sin (\theta )}{4g^{1/2}}\Gamma ^{\hat{r}\hat{%
\phi}}\epsilon +\frac{a^{4}\cos (\theta )}{2r^{4}}\Gamma ^{\hat{\psi}\hat{r}%
}\epsilon +\frac{\cos (\theta )}{4g}\Gamma ^{\hat{\psi}\hat{r}}\epsilon - &&
\notag \\
-\frac{\cos (\theta )}{2}\Gamma ^{\hat{\theta}\hat{\phi}}\epsilon +\frac{%
\cos (\theta )}{4g}\Gamma ^{\hat{\theta}\hat{\phi}}\epsilon  &=&0.
\label{eh4dphi}
\end{eqnarray}%
Through the use of the projection operator (eliminating a further half of
the supersymmetry) 
\begin{equation}
\Gamma ^{\hat{\psi}\hat{r}\hat{\theta}\hat{\phi}}\epsilon =-\epsilon 
\label{projprthph2}
\end{equation}%
it can be shown that all of the terms in equations (\ref{eh4dtheta}) and (%
\ref{eh4dphi}) are satisfied, and so $\epsilon \neq \epsilon (\theta ,\phi )$
, and no Lorentz rotation is necessary for these. Equation (\ref{eh4dpsi})
can be solved with this projection operator giving the Lorentz rotated $%
\epsilon $ 
\begin{equation}
\epsilon =\exp \left\{ -\frac{\psi }{2}\Gamma ^{\hat{\psi}\hat{r}}\right\} 
\tilde{\epsilon}.  \label{eh4epsilon}
\end{equation}

\end{document}